\documentclass{amsart}
\usepackage[dvips]{graphicx}
\theoremstyle{definition}

\theoremstyle{remark}

\numberwithin{equation}{section}
\begin{document}
\keywords{Quantum Field Theory; Higher Order Derivative Theories;
Reality conditions}
\begin{abstract}
The quantization of higher order time derivative  theories including
interactions is unclear. In this paper in order to solve this
problem, we propose to consider a complex  version of the higher
order derivative theory and map this theory to a real first order
theory. To achieve this relationship, the higher order derivative
formulation must be complex since there is not a real canonical
transformation from this theory to a real first order theory with
stable interactions. In this manner, we work with a non-Hermitian
higher order time derivative theory. To quantize this complex
theory, we introduce reality conditions that allow us to map the
complex higher order theory to a real one, and we show that the
resulting theory is regularizable and renormalizable for a class of
interactions.
\end{abstract}
\title{Quantization of the Interacting Non-Hermitian Higher Order Derivative Field}
\author{Carlos A. Margalli \and J. David Vergara}
 \address{ Instituto de Ciencias Nucleares, Universidad Nacional
Aut\'onoma de M\'exico, Apartado Postal 70-543, M\'exico 04510 DF,
M\'exico.}
\email{carlos.margalli@nucleares.unam.mx}
\email{vergara@nucleares.unam.mx}
\date{September 9, 2013}
\maketitle
\section{Introduction}
In physics and mathematics it has been developed some methods for
facing up problems by means of applying an extension of the real
space to the complex plane. In  quantum mechanics, electrodynamics,
quantum field theory and differential equations is common knowledge.
Methods for solving differential equations are narrowly linked to
the higher order derivative mechanics \cite{os} which results in an
interest by encoding the higher order theories to the usual first
order mechanics. These theories have both Lagrangian and Hamiltonian
formulations, and using the latter is, in principle, possible to
quantize the system. The key issue is that it hasn't achieved an
acceptable full quantization of interacting higher order derivative
theories \cite{Smi,Mos}. One always has negative probabilities,
energies unbounded below or a non-unitary dispersion matrix.

Though it appears which the higher order theories aren't a
fundamental theme, many works have showed that using these theories
fundamental problems could be solved. Examples where these theories
arise is F(R) theories, in special the formulation given by Stelle
\cite{stelle} in which it is aggregated a higher order derivative
term that allow to obtain a renormalizable theory, bounded by the
nature of higher order derivative theories.

Other example is the bosonization proposed by Schwinger
\cite{schwinger} for the electrodynamics in 2 dimensions in which
using a non-local transformation it is possible arrive from usual
electrodynamics to the higher order derivative theory with a bosonic
field.

The quantization of the higher order derivative theories isn't a
trivial issue. As early as 1950, Pais and Uhlenbeck established a
non-local transformation that map from a real higher order
derivative theory to the Hamiltonian of two oscillators, with one of
the oscillators with opposite sign in the kinetic term \cite{Pai}. A
subsequent analysis showed that the mapping  described by
Pais-Uhlenbeck point out an inconsistent quantization with problems
as negative probabilities, energy unbounded from below  and a
non-unitary dispersion matrix \cite{Eli}. However, Smilga showed
that if the Pais-Uhlenbeck model is free and  it has different
masses, the inconsistencies don't exist in a quantum theory, but if
masses are equal, Jordan blocks appear implying the loss of
unitarity \cite{Smi}. Subsequent to the Pais-Uhlenbeck model, in
1975  Bernard and  Duncan \cite{berna}, proposed a field theory
model with higher order time derivatives which they try to quantize
using path integrals. Proceeding in this way it was possible to show
that the Matthew's theorem is applicable \cite{berna}. From a model
with different masses Hawking and Hertog proposed that the real
Bernard-Duncan model is set in two independent Hilbert spaces and
resulting that the real higher order derivative theory is acceptable
if it is free \cite{Cer}. In spite of the free Bernard-Duncan model
is quantizable, a way of including interaction potentials is
unfinished still, due to the presence of negative norm states
\cite{Antoniadis}.

The above analysis suggest that the axioms of quantum mechanics
aren't sufficient to establish a consistent quantization for the
higher order derivative theories. In special the Hermiticity axiom
for these theories result incompatible with a higher order
derivative field. Regarding about, a non-Hermitian theory was
proposed by Bender and Manheim \cite{Bender}, who explored this
possibility exploiting the $\mathcal{PT}$-symmetry in order to
determine if a mapping from non-Hermitian theory to Hermitian theory
is possible. For the construction of this non-Hermitian formulation
it is necessary to introduce a new inner product which define a new
$\mathcal{PT}$-quantum mechanics. This suggest the idea of applying
a imaginary scaling transformation that allow to avoid non-Hermitian
$\mathcal{PT}$-symmetric operators \cite{Mos}. Similar to this is to
apply a complex canonical transformation directly \cite{dec}
considering the reality conditions \cite{ashtekar}. In parallel with
this work, it is possible to introduce interactions in the higher
order model using the reality conditions and to develop the complex
structure for higher order derivative mechanics.

The purpose of this work is to show the equivalence between a
complex higher order derivative theory with interactions and a real
first order theory with two scalar fields. The equivalence is
established using reality conditions that cancel the additional
degrees of freedom and map from the complex to the real space.

The higher order derivative theory used as an example is a
complexification of the Bernard-Duncan model \cite{berna}. To start
a quantization by annihilation and creation operators is
established. In that context, using annihilation and creation
operators and the reality conditions, we show the possible
interaction potentials that result in a potential with a stable
critical point.

This paper is organized as follows. Section \ref{operadores}
introduces the key problem of the higher order derivative theories
using the Bernard-Duncan model. After that, we discuss the reality
conditions by means of a simple example given by Ashtekar
\cite{ashtekar}. Section \ref{combedu} presents the complex
Bernard-Duncan theory using higher order derivative fields. These
fields allow to map from a complex theory to a real theory and the
corresponding reality conditions appear into the complex theory. A
Fourier transform let, by means of the higher order derivative
fields, to define annihilation and creation operators. In this part,
the reality conditions are defined in terms of fields. In Section
\ref{crea} we apply the reality conditions in fields by means of
annihilation and creation operators and the commutation relations
between annihilation and creation operators are established. The
higher order derivative Hamiltonian density is found in terms of
these operators using the reality conditions. Finally, we establish
a relation between the complex higher order Hamiltonian theory that
includes the reality conditions and the Hamiltonian theory of two
real Klein-Gordon fields. In Section \ref{interaction}, the
interaction potentials are described so that using the reality
conditions, it is obtained a stable interaction with a critic point
that allows to do a perturbative expansion and we show that the
resulting theory is regularizable and renormalizable. Finally, in
Section \ref{conclu} we summarize our results.

\section{Creation and Annihilation Operators in the real theory}\label{operadores}

In order to analyze problems that appear when we quantize a higher
order temporal derivative theory, we introduce the Bernard-Duncan
model \cite{berna} by means of its real Lagrangian density
\begin{equation}\label{paisimple}
 \mathcal{L}_{BD}=-\frac{1}{2}(\partial_{\mu}\partial^{\mu}\varphi)^{2}+\frac{(m_{1}^{2}
+m_{2}^{2})}{2}\partial_{\mu}\varphi\partial^{\mu}\varphi-\frac{m_{1}^{2}m_{2}^{2}}{2}
\varphi^{2},
\end{equation}
which generates the equation of motion
\begin{equation}\label{ecuacionbd}
 \Box^{2}\varphi+(m_{1}^{2}+m_{2}^{2})\Box\varphi+m_{1}^{2}m_{2}^{2}\varphi=0.
\end{equation}
Using the Lagrangian density (\ref{paisimple}) and the Ostrogradsky theory \cite{os}, we
obtain the higher order derivative momenta for the fields $\varphi$ and $\dot{\varphi}$
\begin{eqnarray}\label{momentosbdr}
 \pi_{\dot{\varphi}}=-\Box\varphi,\\
\pi_{\varphi}=\varphi^{(3)}-\nabla^{2}\frac{d}{dt}\varphi+(m_{1}^{2}+m_{2}^{2})
\frac{d}{dt}\varphi. \nonumber
\end{eqnarray}
The above equations will allow to define a symplectic structure of
the phase space considering that the real Lagrangian depends on
$(\varphi,\dot \varphi, \ddot \varphi)$.

Using the Fourier transform
\begin{equation}\label{Fourier}
 \varphi(\vec{x},t)=\int\!\frac{d^{3}p}{(2\pi)^{\frac{3}{2}}}e^{i\vec{p}\cdotp \vec{x}}\Psi(\vec{p},t),
\end{equation}
the equation of motion (\ref{ecuacionbd}) results
\begin{equation}\label{energiabd}
\Psi^{(4)}(\vec{p},t)+(E_{1}^{2}+E_{2}^{2})\Psi^{(2)}(\vec{p},t)+E_{1}^{2}E_{2}^{2}\Psi(\vec{p},t)=0.
 \end{equation}
The general solution to (\ref{energiabd}) is
\begin{equation}
\Psi(\vec{p},t)=\textbf{\itshape {a}}(\vec{p})e^{-iE_{1}t}+\textbf{\itshape {c}}(-\vec{p})e^{iE_{1}t}
+\textbf{\itshape {b}}(\vec{p})e^{-iE_{2}t}+\textbf{\itshape {d}}(-\vec{p})e^{iE_{2}t}.
\end{equation}
In the standard formalism is requested that the Lagrangian density
to be real. In consequence, the field $\varphi(\vec{x},t)$ must be
real which imposes a restriction in the Fourier coefficients
\begin{equation}\label{realidadcon}
 \textbf{\itshape {d}}=\textbf{\itshape {b}}^{*},\qquad \textbf{\itshape {c}}=\textbf{\itshape {a}}^{*}.
\end{equation}
With the real solution of the field for (\ref{ecuacionbd}), we
obtain hermiticity when a quantization is done by means of promote
the Fourier coefficients to operators.

The solution which include (\ref{realidadcon}), which obey
(\ref{energiabd}) and which induce a real $\varphi$ in
(\ref{Fourier}) is
\begin{equation}
\Psi(\vec{p},t)=\textbf{\itshape {a}}(\vec{p})e^{-iE_{1}t}+\textbf{\itshape {a}}^{*}(-\vec{p})e^{iE_{1}t}
+\textbf{\itshape {b}}(\vec{p})e^{-iE_{2}t}+\textbf{\itshape {b}}^{*}(-\vec{p})e^{iE_{2}t}.
\end{equation}
In order to quantize the system  and following the usual rules, we
promote coefficients $\textbf{\itshape {a}}$ and $\textbf{\itshape
{b}}$ to operators, the general solution to (\ref{ecuacionbd}) is
\begin{eqnarray}\label{cuanti}
\varphi(\vec{x},t)=\int\frac{d^{3}p}{(2\pi)^{\frac{3}{2}}}
\left\lbrace
\frac{1}{(2E_{1})^{\frac{1}{2}}(m_{2}^{2}-m_{1}^{2})^{\frac{1}{2}}}[\textbf{\itshape
{a}} (\vec{p})e^{i\vec{p}\cdotp\vec{x}-iE_{1}t}+
\textbf{\itshape {a}}^{\dagger}(\vec{p})e^{-i\vec{p}\cdotp \vec{x}+iE_{1}t}]\right. \nonumber\\
\left.
+\frac{1}{(2E_{2})^{\frac{1}{2}}(m_{2}^{2}-m_{1}^{2})^{\frac{1}{2}}}[\textbf{\itshape
{b}} (\vec{p})e^{i\vec{p}\cdotp\vec{x}-iE_{2}t} +\textbf{\itshape
{b}}^{\dagger}(\vec{p})e^{-i\vec{p}\cdotp\vec{x}+iE_{2}t}]\right\rbrace
,
\end{eqnarray}
and from this expression, we obtain the reality condition
$\varphi=\varphi^{\dagger}$, which implies that the field $\varphi$
is hermitic. The solution (\ref{cuanti}) is Lorentz invariant and it
makes sense only in the case $m_{2}\neq m_{1}$, along this article
we will use this condition.

The case $m_{1}= m_{2}$ has been analyzed in \cite{Bender} and we
can use similar arguments. However, from (\ref{cuanti}) is possible
to find $\dot{\varphi}$ that is a field in the Ostrogradsky's theory
and to obtain the momenta (\ref{momentosbdr}) in terms of
annihilation and creation operators.

The commutators associated to the annihilation and the creation
operators resulting from the canonical commutators are
\begin{eqnarray}
 [\textbf{\itshape {a}}(\vec{p}),\textbf{\itshape {a}}^{\dagger}(\vec{p'})]=
\delta(\vec{p}-\vec{p'}),\\
\
[\textbf{\itshape {b}}(\vec{p}),\textbf{\itshape {b}}^{\dagger}(\vec{p'})]=-\delta(\vec{p}-\vec{p'}).\label{calbd}
\end{eqnarray}
The sign in the commutation relation (\ref{calbd}) is the root of the problem to
quantize the higher order derivative theories.

For example, considering  the higher order derivative theory in
(\ref{paisimple}), we get the Hamiltonian density by means of the
Ostrogradsky method
\begin{equation}\label{ostrogra}
 \mathcal{H}_{BD}=\pi_{\varphi}\frac{d\varphi}{dt}+\pi_{\dot{\varphi}}\frac{d^{2}\varphi}{dt^{2}}-\mathcal{L}_{BD},
\end{equation}
with the Hamiltonian density given by
\begin{equation}\label{hamiltonparcial}
 \mathcal{H}_{BD}=\pi_{\varphi}\dot{\varphi}-\frac{1}{2}\pi^{2}_{\dot{\varphi}}-\frac{(m_{1}^{2}
 +m_{2}^{2})}{2}\dot{\varphi}^{2}+\frac{m_{1}^{2}m_{2}^{2}}{2}\varphi^{2}
 +\pi_{\dot{\varphi}}\nabla^{2}\varphi+ \frac{(m_{1}^{2}+m_{2}^{2})}{2}(\nabla\varphi)^{2}
\end{equation}
and with the respective phase space
$(\varphi,\pi_{\varphi},\dot{\varphi},\pi_{\dot{\varphi}})$.

In terms of annihilation and creation operators the Hamiltonian
density  (\ref{hamiltonparcial}) that is  obtained by Ostrogradsky's
method (\ref{ostrogra}) is unbounded from below and results
\begin{equation}\label{paisuhle}
 \ \ H_{BD}=\int d^{3}p  \left\lbrace  \frac{E_{1}}{2}[\textbf{\itshape {a}}^{\dagger}
 (\vec{p})\textbf{\itshape {a}}(\vec{p})+\textbf{\itshape {a}}(\vec{p})\textbf{\itshape {a}}^{\dagger}
 (\vec{p})]-\frac{E_{2}}{2} [\textbf{\itshape {b}}^{\dagger}(\vec{p})\textbf{\itshape {b}}(\vec{p})
+\textbf{\itshape {b}}(\vec{p})\textbf{\itshape {b}}^{\dagger}(\vec{p})]\right\rbrace.
\end{equation}
The commutators (\ref{calbd}) generate negative norm states or
negative probabilities, so this theory isn't a good quantum theory.
Because, there isn't an interaction potential here and the free
system doesn't interchange energy from one field to another field,
so it is correct to think that the system can be divided in two
independent free systems \cite{Cer}. However, the self-energy
contribution manifest an internal interaction in the system which is
induced by an external agent so consequently, a system without a
self-interaction potential is a non-physical system. To establish in
the Bernard-Duncan model an interaction potential, that can be
handled using perturbative theory, using the approach
(\ref{paisuhle}) is impossible. For that reason, we think that is
necessary to relax the Hermiticity condition for $\varphi$ that is
to say $\varphi(x)\neq\varphi^{\dagger}$. The idea of a reality
conditions exposed by Ashtekar in the case of gravitation
\cite{ashtekar} is to replace the Hermiticity condition in order to
set a new Hermiticity condition least restrictive which allows a
complex higher order derivative field and a possible solution to the
problem in (\ref{paisuhle}). In the next subsection, we will review
this strategy.

\subsection{Reality Conditions}\label{conreashtekar}
The complexification by means of an extended space is a traditional
method in mathematics and physics which is used to solve several
problems in different branches of the science.  In our case, we
don't focus in the complexification, but we focus in reality
conditions which allow to reduce and to solve a problem by means of
projecting to the real space. It is possible to understand the
complexification as an extension of the physical degrees of freedom
and after that to build  a projection from the complex to the real
space. Bender proposed a similar situation \cite{Bender}, changing
the internal product using the PT symmetry as an assistance to find
the correct internal product. In our case the reality conditions
will provide the appropriate projection and also the internal
product. To introduce this proposal, we consider a simple example
that allows to illustrate some consequences of using this method.

Let us consider the harmonic oscillator with phase space
$\Gamma=(q,p)$ in two dimensions and a real Hamiltonian
\begin{equation}\label{ashtekar}
 h(q,p)=\frac{1}{2}(q^{2}+p^{2}).
\end{equation}
Now, we want to extend the domain of definition to the complex space, then a new variable is used
\begin{equation}\label{defico}
z\equiv q-ip,
\end{equation}
where the pair $\tilde{\Gamma}=(q,z)$, is the new complex phase
space, with Poisson brackets defined by
\begin{eqnarray}\label{PBz}
\left\lbrace q,q  \right\rbrace=0,\qquad \left\lbrace z,z
\right\rbrace=0,\qquad \left\lbrace z,q \right\rbrace =i,
\end{eqnarray}
which will be thought as a canonical conjugate set. Introducing a
function $f(q,p)$ on $\Gamma$ is possible to define a new function
on $\tilde{\Gamma}$ using (\ref{defico})
\begin{equation}
g(q,z)\equiv f(q,i(z-q))
\end{equation}
and any function can be constructed in this way, taking care of
computing the evolution by means of the Poisson brackets
(\ref{PBz}).

In particular the Hamiltonian function in terms of $(q,z)$ is
\begin{equation}\label{complex}
h(q,z)=\frac{1}{2}(q^{2}-(z-q)^{2})=zq-\frac{1}{2}z^{2}
\end{equation}
and using commutators the temporal evolution is
\begin{equation}\label{evtem}
\dot{q}=\left\lbrace q,h \right\rbrace=iz-iq, \qquad
\dot{z}=\left\lbrace z,h\right\rbrace=iz.
\end{equation}
The equations (\ref{evtem}) are equations of motion for the complex
phase space $\tilde{\Gamma}$. However, this description have to be
consistent with the equations of motion resulting from
(\ref{ashtekar}) in order to preserve the original dynamics. Now, to
project from the complex space $(q,z)$ to the original real space we
propose the following reality conditions
\begin{equation}\label{codere}
 q=q^{*},\qquad z^{*}=(-z+2q).
\end{equation}
The first equation in (\ref{codere}) set a real $q$ and it is similar to the initial real phase space. On the other hand
the second equation recovers the initial real phase space preserving $p=p^{*}$.
This conditions impose a constraint for $z$ in the new complex phase space.

To put it briefly, we have first a mapping from a real phase space
to a complex space and in order to make a consistent theory and a
compatible dynamics, we introduce the reality conditions
(\ref{codere}) on $\tilde{\Gamma}$.

Note that the relationship (\ref{defico}) obeys the reality conditions and provide a direct mapping
which carries from the equations of motion (\ref{evtem}) to the equations of motion generated by (\ref{ashtekar}).
The above idea will be applied in the Complex Bernard Duncan model.

\section{Complex Bernard-Duncan Theory and Rea\-lity conditions}\label{combedu}
There are a lot of higher order derivative field theories, but in
order to understand their characteristics, we choose the most
simplest the Bernard-Duncan model.

In this section it will be analyzed the consequence of extending the
field theory (\ref{paisimple}) to the complex plane, i.e. we
consider that the field $\phi$ is defined by
\begin{equation}\label{separado}
 \phi\equiv\phi_{R}+i\phi_{I},
\end{equation}
with the complex higher order derivative action given by
\begin{equation}\label{acsch}
 W=\int\!d^{4}x \frac{1}{2}[-(\Box \phi)^{2}+(m_{1}^{2}+m_{2}^{2})
 \partial_{\mu}\phi\partial^{\mu}\phi-m_{1}^{2}m_{2}^{2}\phi^{2}].
\end{equation}
From the expression (\ref{acsch}), it is possible to make a
variation in $\phi$ and we obtain
\begin{eqnarray}\label{caco}
 \delta W=\int\!d^{4}x -\frac{\partial^{2}}{\partial t^{2}}(\Box\phi)\delta\phi+
 (\Box\phi)\nabla^{2}\delta\phi-(m_{1}^{2}+m_{2}^{2})\frac{d^{2}}{dt^{2}}\phi\delta\phi\\
\nonumber -(m_{1}^{2}+m_{2}^{2})\nabla\phi\cdotp\delta\nabla\phi
-m_{1}^{2}m_{2}^{2}\phi\delta\phi\\\nonumber +\int\!d^{3}x
-\Box\phi\delta\frac{d}{dt}\phi+(m_{1}^{2}+m_{2}^{2})\frac{d}{dt}\phi\delta\phi-\frac{d}{dt}(-\Box\phi)\delta\phi.\nonumber
\end{eqnarray}
From the last expression, we obtain the momenta
\begin{eqnarray}
\label{momentospu}
 \pi_{\dot{\phi}}=-\Box\phi,\\
\pi_{\phi}=\phi^{(3)}-\nabla^{2}\frac{d}{dt}\phi+(m_{1}^{2}+m_{2}^{2})\frac{d}{dt}\phi
\nonumber
\end{eqnarray}
then the variation of the action can be summarized as
\begin{equation}\label{eqofmo}
 \delta W=\int\!d^{4}x-[\Box^{2}\phi+(m_{1}^{2}+m_{2}^{2})\Box\phi+m_{1}^{2}m_{2}^{2}\phi]\delta\phi+\int\!d^{3}x[\pi_{\dot{\phi}}\delta\dot{\phi}+\pi_{\phi}\delta\phi].
\end{equation}
The complex equation of motion can be identified from (\ref{eqofmo})
\begin{equation}\label{ecuacion}
 \Box^{2}\phi+(m_{1}^{2}+m_{2}^{2})\Box\phi+m_{1}^{2}m_{2}^{2}\phi=0
\end{equation}
and using the Fourier transformation (\ref{Fourier}) applied to this
case, we obtain
\begin{equation}\label{energia}
\psi^{(4)}(\vec{p},t)+(E_{1}^{2}+E_{2}^{2})\psi^{(2)}(\vec{p},t)+E_{1}^{2}E_{2}^{2}\psi(\vec{p},t)=0,
 \end{equation}
where $E_{1}^{2}=(\vec{p}^{2}+m_{1}^{2})$ and
$E_{2}^{2}=(\vec{p}^{2}+m_{2}^{2})$ are two energies with different masses.
In this case the field $\phi$ is complex then to determine $\psi$, we use a new
set of reality conditions.

These conditions will cancel the additional degrees of freedom that
appear from the complexification of the system. The equation
(\ref{energia}) can be rewritten as
\begin{equation}\label{energiados}
 \left(\frac{d^{2}}{dt^{2}}+E_{1}^{2} \right) \left( \frac{d^{2}}{dt^{2}}+E_{2}^{2} \right)\psi(\vec{p},t)=0
\end{equation}
and the general solution is
\begin{equation}
 \psi(\vec{p},t)=\textbf{f}(\vec{p})e^{-iE_{1}t}+\textbf{c}(-\vec{p})e^{iE_{1}t}+\textbf{b}(\vec{p})e^{-iE_{2}t}+\textbf{d}(-\vec{p})e^{iE_{2}t}.
\end{equation}
In particular, we look for a Lorentz invariant solution to (\ref{ecuacion}) which is
\begin{eqnarray}\label{phicreauno}
\phi(\vec{x},t)=\int\!\frac{d^{3}p}{(2\pi)^{\frac{3}{2}}} \frac{1}{(2E_{1})^{\frac{1}{2}}}[\textbf{f}(\vec{p})e^{i\vec{p}\cdotp\vec{x}-iE_{1}t}+\textbf{c}
(\vec{p})e^{-i\vec{p}\cdotp \vec{x}+iE_{1}t}]
\\
+\frac{1}{(2E_{2})^{\frac{1}{2}}}[\textbf{b}(\vec{p})e^{i\vec{p}\cdotp\vec{x}-iE_{2}t}
+\textbf{d}(\vec{p})e^{-i\vec{p}\cdotp\vec{x}+iE_{2}t}]\nonumber
\end{eqnarray}
and using this field, we obtain $\dot{\phi}$ and the corresponding
momenta in the Ostrogradsky formulation $\pi_{\phi}$,
$\pi_{\dot{\phi}}$ that impose a relationship between the complex
fields and momenta with the annihilation and creation operators.

With the new fields and momenta (\ref{momentospu}) the resulting
Hamiltonian density is
\begin{eqnarray}\label{cons}
\mathcal{H}_{BD}=\pi_{\phi}\dot{\phi}-\frac{(\pi_{\dot{\phi}})^{2}}{2}-\frac{(m_{1}^{2}+m_{2}^{2})}{2}\dot{\phi}^{2}
+\frac{(m_{1}^{2}+m_{2}^{2})}{2}(\nabla\phi)^{2}+\frac{m_{1}^{2}m_{2}^{2}}{2}\phi^{2}\nonumber\\
-\nabla\phi\cdotp\nabla(\pi_{\dot{\phi}}).
\end{eqnarray}
The Hamiltonian density (\ref{cons}) is similar to (\ref{hamiltonparcial}), but (\ref{cons})
is complex while the density (\ref{hamiltonparcial}) is real and different from (\ref{cons}) by  a total spatial derivative.

Following the idea of Section \ref{conreashtekar}, we introduce a
mapping from the complex to the real space. To define this mapping,
we implement a canonical transformation between the complex phase
space $(\phi, \dot{\phi}, \pi_{\phi},\pi_{\dot{\phi}})$ to the real
phase space $(\psi_{1},\pi_{\psi_{1}},\psi_{2},\pi_{\psi_{2}})$
defined as
\begin{eqnarray}\label{realidad}
\psi_{1}=\frac{1}{(m_{1}^{2}-m_{2}^{2})^{\frac{1}{2}}}(im_{2}^{2}\phi
-i\pi_{\dot{\phi}}),\qquad \psi_{2}
=\frac{1}{(m_{1}^{2}-m_{2}^{2})^{\frac{1}{2}}}(m_{1}^{2}\phi-\pi_{\dot{\phi}}),\\
\pi_{\psi_{1}}=i\frac{1}{(m_{1}^{2}-m_{2}^{2})^{\frac{1}{2}}}(\pi_{\phi}-m_{1}^{2}\dot{\phi}),
\qquad  \pi_{\psi_{2}}=\frac{1}{(m_{1}^{2}-m_{2}^{2})^{\frac{1}{2}}}(\pi_{\phi}-m_{2}^{2}\dot{\phi})
\nonumber
\end{eqnarray}
which is a local and linear canonical transformation. In order to fully establish that the
fields $\psi_{1}$ and $\psi_{2}$ and the respective momenta
$\pi_{\psi_{1}}, \pi_{\psi_{2}}$, are real we assume that are
complex fields and  impose that the imaginary parts are zero.

The real and imaginary parts of the complex fields and the complex momenta are
\begin{eqnarray}\label{conrepa}
(\psi_{1R}+i\psi_{1I})=\frac{1}{(m_{1}^{2}-m_{2}^{2})^{\frac{1}{2}}}[(\pi_{\dot{\phi}_{I}}-m_{2}^{2}\phi_{I})-i(\pi_{\dot{\phi}_{R}}-m_{2}^{2}\phi_{R})],\nonumber\\
(\psi_{2R}+i\psi_{2I})=\frac{1}{(m_{1}^{2}-m_{2}^{2})^{\frac{1}{2}}}[(m_{1}^{2}\phi_{R}
-\pi_{\dot{\phi}_{R}})
-i(\pi_{\dot{\phi}_{I}}-m_{1}^{2}\phi_{I})],\nonumber\\
(\pi_{\psi_{1R}}+i\pi_{\psi_{1I}})=\frac{1}{(m_{1}^{2}-m_{2}^{2})^{\frac{1}{2}}}[-(\pi_{\phi_{I}}-m_{1}^{2}\dot{\phi}_{I})
+i(\pi_{\phi_{R}}-m_{1}^{2}\dot{\phi}_{R})],\nonumber\\
(\pi_{\psi_{2R}}+i\pi_{\psi_{2I}})=\frac{1}{(m_{1}^{2}-m_{2}^{2})^{\frac{1}{2}}}[(\pi_{\phi_{R}}-m_{2}^{2}\dot{\phi}_{R})
+i(\pi_{\phi_{I}}-m_{2}^{2}\dot{\phi}_{I})].
\end{eqnarray}
Therefore, the equations (\ref{conrepa}) impose 4 conditions
\begin{eqnarray}\label{reali}
 (\pi_{\dot{\phi}_{R}}-m_{2}^{2}\phi_{R})=0,\qquad (\pi_{\dot{\phi}_{I}}-m_{1}^{2}\phi_{I})=0,\\
(\pi_{\phi_{R}}-m_{1}^{2}\dot{\phi}_{R})=0,\qquad
(\pi_{\phi_{I}}-m_{2}^{2}\dot{\phi}_{I})=0,\nonumber
\end{eqnarray}
these are the reality conditions for the complex fields $(\phi,\dot{\phi},\pi_{\phi},\pi_{\dot{\phi}})$.

Using these conditions, the relationship between components of the complex fields and  the
fields $(\psi_{1},\psi_{2},\pi_{\psi_{1}},\pi_{\psi_{2}})$ are
\begin{eqnarray}\label{contact}
 \psi_{2}=(m_{1}^{2}-m_{2}^{2})^{\frac{1}{2}}\phi_{R},\qquad \psi_{1}=(m_{1}^{2}-m_{2}^{2})^{\frac{1}{2}}\phi_{I},\\
\pi_{\psi_{2}}=(m_{1}^{2}-m_{2}^{2})^{\frac{1}{2}}\dot{\phi}_{R},\qquad \pi_{\psi_{1}}=(m_{1}^{2}-m_{2}^{2})^{\frac{1}{2}}\dot{\phi}_{I}.
\end{eqnarray}
In addition to the above issues we work in the Hamiltonian
formulation because  it is easy to define a complex canonical
transformation (\ref{realidad}) instead of a non local complex
transformation that is defined in the Lagrangian formulation
\cite{dec}. In terms of the Lagrangian formulation, the non local
complex transformation point out troubles as linearity, and
simultaneity. However we wish to remark by means of the Hamiltonian
formulation that it is possible to introduce a consistent theory
including a Lagrangian formulation using a Legendre transformation.

In (\ref{conrepa}), we impose that the fields $(\psi_{1},\pi_{\psi_{2}},\psi_{2},\pi_{\psi_{2}},)$ are
real, these conditions constraint
the complex higher order fields $(\phi,\pi_{\phi}, \dot{\phi}, \pi_{\dot{\phi}})$, then we
want to rewrite these conditions in terms of the higher order fields and their complex
conjugate fields so we get the expressions
\begin{eqnarray}\label{realidaduno}
 (m_{1}^{2}-m_{2}^{2})\phi^{*}=(m_{1}^{2}+m_{2}^{2})\phi-2\pi_{\dot{\phi}}\nonumber\\ (m_{1}^{2}-m_{2}^{2})\dot{\phi}^{*}=-(m_{1}^{2}+m_{2}^{2})\dot{\phi}+2\pi_{\phi}\nonumber\\
(m_{1}^{2}-m_{2}^{2})\pi^{*}_{\phi}=(m_{1}^{2}+m_{2}^{2})\pi_{\phi}-2m_{1}^{2}m_{2}^{2}\dot{\phi}\nonumber\\  (m_{1}^{2}-m_{2}^{2})\pi^{*}_{\dot{\phi}}=-(m_{1}^{2}+m_{2}^{2})\pi_{\dot{\phi}}+2m_{1}^{2}m_{2}^{2}\phi.
\end{eqnarray}
In this way, starting with a complex higher order theory, using a canonical transformation
and applying the reality conditions we were able to reduce the complex theory to a real one.

Now, we need to figure out the nature of the dynamics of the new fields \break
 $(\psi_{1},\psi_{2},\pi_{\psi_{1}},\pi_{\psi_{2}})$. To do that we compute the Hamiltonian density (\ref{cons}) in terms of these fields and we get
\begin{equation}\label{hamiltonkg}
 \mathcal{H}_{KG}=\frac{\pi^{2}_{\psi_{1}}}{2}+\frac{(\nabla\psi_{1})^{2}}{2}+\frac{m_{1}^{2}\psi_{1}^{2}}{2}
+\frac{\pi^{2}_{\psi_{2}}}{2}+\frac{(\nabla\psi_{2})^{2}}{2}+\frac{m_{2}^{2}\psi_{2}^{2}}{2}.
\end{equation}
This Hamiltonian density corresponds to the Hamiltonian density of
two Klein-Gordon fields and given that the fields
$(\psi_{1},\psi_{2},\pi_{\psi_{1}},\pi_{\psi_{2}})$ are real
quantities, we finish with an ordinary first order theory. So,
summarizing by applying a canonical transformation and the reality
conditions we were able to map a complex higher order derivative
theory to a real first order derivative theory. So the reality
conditions reduce degrees of freedom of the complex higher order
theory from eight to four per point and these conditions can be
interpreted as second class constraints in the Dirac's formalism \cite{Margalli.Vergara.2}.

In the next Section, we shall describe how to quantize the complex Bernard-Duncan model  applying
reality conditions (\ref{realidaduno}) on the creation and annihilation operators.

\section{The Reality Conditions using Creation and Annihilation operators}\label{crea}
In order to quantize any system, it is usual to promote fields and momenta to operators and
if these are real quantities they acquire Hermiticity properties. In this section, we build
a quantum theory without using the Hermiticity axiom and using instead the reality conditions.

As a starting point, we work with a complex higher order time
derivative theory and a complex higher order field which is not
Hermitic, but it satisfy the reality conditions. In the quantization
this property is inherited and we assume it. Using the reality
conditions (\ref{realidaduno}) and the fields and momenta resulting
from (\ref{phicreauno}), we want to determine the reality conditions
in terms of creation and annihilation operators then, the conditions
are
\begin{eqnarray}\label{corean}
 \textbf{c}^{*}=-\textbf{f},\qquad \textbf{d}^{*}=\textbf{b}.
\end{eqnarray}
The reality conditions in terms of this Fourier coefficients
(\ref{corean}) differ from the conditions (\ref{realidadcon})
because we use at the beginning that the fields $(\phi,
\dot{\phi},\pi_{\phi},\pi_{\dot{\phi}})$ are complex.

Using the reality conditions (\ref{corean}) into the field (\ref{phicreauno}), we get
\begin{eqnarray}\label{phi}
 \phi(\vec{x},t)=\int\!\frac{d^{3}p}{(2\pi)^{\frac{3}{2}}} \left\lbrace \frac{1}{(2E_{1})^{\frac{1}{2}}}[\hat{\textbf{f}}(\vec{p})e^{i\vec{p}\cdotp\vec{x}-iE_{1}t}-\hat{\textbf{f}}^{\ \dagger}
(\vec{p})e^{-i\vec{p}\cdotp \vec{x}+iE_{1}t}]\right. \\
\left. +\frac{1}{(2E_{2})^{\frac{1}{2}}}[\hat{\textbf{b}}(\vec{p})e^{i\vec{p}\cdotp\vec{x}-iE_{2}t}
+\hat{\textbf{b}}^{\dagger}(\vec{p})e^{-i\vec{p}\cdotp\vec{x}+iE_{2}t}]\right\rbrace .\nonumber
\end{eqnarray}
In fact any operator can be written in terms of a Hermitian part together with the anti-Hermitian part. Thus
\begin{equation}\label{heran}
 \hat{\phi}=\hat{\phi}^{f}_{A}+\hat{\phi}^{b}_{H}
\end{equation}
where it must be  emphasized that there is a Hermitian part $\hat{\phi}^{b}_{H}$  that is originated by $\hat{\textbf{b}}$ and there is an anti-Hermitian part $\hat{\phi}^{f}_{A}$ that is given by $\hat{\textbf{f}}$.

However, we can use the property of anti-Hermitian operators that is
\begin{equation}
 \hat{\mathcal{O}}_{A}=i\hat{\mathcal{O}}_{H},
\end{equation}
where an anti-Hermitian operator is written as $i$ times a Hermitian operator. Using this property in (\ref{heran}) for
$\hat{\textbf{f}}$, we obtain
\begin{equation}\label{anhe}
\hat{\textbf{f}}=i\hat{\textbf{a}}.
\end{equation}
This expression clarify the meaning of operator $\hat{\textbf{f}}$
that is an annihilation operator that can be used to build an
anti-Hermitian operator.

Now, using (\ref{anhe}) in the higher order field (\ref{phi}) in
order to include this property and to build a theory with Hermitian
operators, we obtain
\begin{eqnarray}\label{phidos}
 \phi(\vec{x},t)=\int\!\frac{d^{3}p}{(2\pi)^{\frac{3}{2}}}( \frac{1}{(2E_{1})^{\frac{1}{2}}}[i\hat{\textbf{a}}(\vec{p})
e^{i\vec{p}\cdotp\vec{x}-iE_{1}t}+i\hat{\textbf{a}}^{\dagger}(\vec{p})e^{-i\vec{p}\cdotp \vec{x}+iE_{1}t}]
\\
+\frac{1}{(2E_{2})^{\frac{1}{2}}}[\hat{\textbf{b}}(\vec{p})e^{i\vec{p}\cdotp\vec{x}-iE_{2}t}
+\hat{\textbf{b}}^{\dagger}(\vec{p})
e^{-i\vec{p}\cdotp\vec{x}+iE_{2}t}])\nonumber
\end{eqnarray}
looking at this expression, we see that the field $\phi(\vec{x},t)$ isn't real.

By applying the reality conditions, we can separate the imaginary and real parts, that are
given by
\begin{eqnarray}\label{realimaginaria}
\phi_{R}=\int\!\frac{d^{3}p}{(2\pi)^{\frac{3}{2}}}\frac{1}{(2E_{2})^{\frac{1}{2}}}[\hat{\textbf{b}}(\vec{p})e^{i\vec{p}\cdotp\vec{x}-iE_{2}t}
+\hat{\textbf{b}}^{\dagger}(\vec{p})
e^{-i\vec{p}\cdotp\vec{x}+iE_{2}t}],\\\nonumber
\phi_{I}=\int\!\frac{d^{3}p}{(2\pi)^{\frac{3}{2}}}\frac{1}{(2E_{1})^{\frac{1}{2}}}[\hat{\textbf{a}}(\vec{p})
e^{i\vec{p}\cdotp\vec{x}-iE_{1}t}+\hat{\textbf{a}}^{\dagger}(\vec{p})e^{-i\vec{p}\cdotp \vec{x}+iE_{1}t}]\nonumber
\end{eqnarray}
considering that $\phi_{R}$ and $\phi_{I}$ are Hermitian independent fields.

From (\ref{phidos}), higher order fields and momenta are
\begin{eqnarray}
\dot{\phi}(\vec{x},t)=\int\!\frac{d^{3}p}{(2\pi)^{\frac{3}{2}}} \frac{1}{(2E_{1})^{\frac{1}{2}}}[i(-iE_{1})\hat{\textbf{a}}(\vec{p})
e^{i\vec{p}\cdotp\vec{x}-iE_{1}t}+i(iE_{1})\hat{\textbf{a}}^{\dagger}(\vec{p})e^{-i\vec{p}\cdotp \vec{x}+iE_{1}t}]
\nonumber\\
+\frac{1}{(2E_{2})^{\frac{1}{2}}}[(-iE_{2})\hat{\textbf{b}}(\vec{p})e^{i\vec{p}\cdotp\vec{x}-iE_{2}t}+(iE_{2})\hat{\textbf{b}}^{\dagger}(\vec{p})e^{-i\vec{p}\cdotp\vec{x}+iE_{2}t}],
\\
 \pi_{\phi}(\vec{x},t)=\int\!\frac{d^{3}p}{(2\pi)^{\frac{3}{2}}} \frac{iE_{1}m_{2}^{2}}{(2E_{1}^{\frac{1}{2}})}
[-i\hat{\textbf{a}}(\vec{p})e^{i\vec{p}\cdotp\vec{x}-iE_{1}t}+i\hat{\textbf{a}}^{\dagger}(\vec{p})
e^{-i\vec{p}\cdotp \vec{x}+iE_{1}t}]\nonumber\\
+\frac{iE_{2}m_{1}^{2}}{2E_{2}^{\frac{1}{2}}}[-\hat{\textbf{b}}(\vec{p})e^{i\vec{p}\cdotp\vec{x}-iE_{2}t}+\hat{\textbf{b}}^{\dagger}(\vec{p})
e^{-i\vec{p}\cdotp\vec{x}+iE_{2}t}],\\
\pi_{\dot{\phi}}(\vec{x},t)=\int\!\frac{d^{3}p}{(2\pi)^{\frac{3}{2}}} \frac{m_{1}^{2}}{(2E_{1}^{\frac{1}{2}})}
[i\hat{\textbf{a}}(\vec{p})e^{i\vec{p}\cdotp\vec{x}-iE_{1}t}+i\hat{\textbf{a}}^{\dagger}(\vec{p})e^{-i\vec{p}\cdotp \vec{x}+iE_{1}t}]
\nonumber\\
+\frac{m_{2}^{2}}{(2E_{2})^{\frac{1}{2}}}[\hat{\textbf{b}}(\vec{p})e^{i\vec{p}\cdotp\vec{x}-iE_{2}t}
+\hat{\textbf{b}}^{\dagger}(\vec{p})
e^{-i\vec{p}\cdotp\vec{x}+iE_{2}t}],\label{momentopiuno}
\end{eqnarray}
where it is important emphasize that these aren't Hermitian
quantities. Knowing the method that imply to use reality conditions
and the quantization rules that are applied for annihilation and
creation operators, we will be able to face troubles as negative
norm states or energy unbounded from below. In the next section, we will
calculate commutators for annihilation and creation operators using
the tools that here we have described. This in the future will allow
us to calculate the energy without any problem and to obtain quantum
states with positive probability.

\subsection{Commutation Relations between Creation and Annihilation Operators}
The key problem in higher order time derivative theories are the
commutators. From the equation (\ref{calbd}) is possible to find
negative norm states resulting from the Hermitian condition. However
this problem is faced using the reality conditions, because we
achieve that the wrong sign in (\ref{calbd}) disappears and we
obtain positive probabilities.

In order to show the effect of reality conditions, we consider the
commutators for the conjugate variables in the complex theory. We
establish that the parenthesis for higher order fields and momenta
obey  usual expressions given by
\begin{equation}\label{poisson}
 \{\phi(t,\vec{x}), \pi_{\phi}(t,\vec{x}_{0})\}=\delta(\vec{x}-\vec{x}_{0}) ,
\qquad \{\dot{\phi}(t,\vec{x}), \pi_{\dot{\phi}}(t,\vec{x}_{0})\}=\delta(\vec{x}-\vec{x}_{0}).
\end{equation}
From these classical expressions, we can promote fields and momenta
to operators and we determine commutators for annihilation and
creation operators. Using the full expressions in  (\ref{poisson}),
we  have
\begin{equation}\label{joel}
 [\hat{\textbf{a}}(\vec{p}),\hat{\textbf{a}}^{\dagger}(\vec{p}_{0})]=f(\vec{p})\delta(\vec{p}-\vec{p}_{0}),\qquad
 [\hat{\textbf{b}}(\vec{p}),\hat{\textbf{b}}^{\dagger}(\vec{p}_{0})]=g(\vec{p})\delta(\vec{p}-\vec{p}_{0}).
\end{equation}
 From (\ref{joel}) on (\ref{poisson}), we obtain two conditions
\begin{equation}
 -\frac{1}{2}m_{2}^{2}f(\vec{p})+\frac{1}{2}m_{1}^{2}g(\vec{p})=\frac{1}{2},\qquad
 \frac{1}{2}m_{1}^{2}f(\vec{p})-\frac{1}{2}m_{2}^{2}g(\vec{p})=\frac{1}{2},
\end{equation}
resulting in
\begin{equation}
 f(\vec{p})=\frac{1}{(m_{1}^{2}-m_{2}^{2})},\qquad g(\vec{p})=\frac{1}{(m_{1}^{2}-m_{2}^{2})}.
\end{equation}
In consequence the commutators are
\begin{eqnarray}\label{conmutacion}
 [\hat{\textbf{a}}(\vec{p}),\hat{\textbf{a}}^{\dagger}(\vec{p}_{0})]=\frac{1}{(m_{1}^{2}-m_{2}^{2})}\delta(\vec{p}
-\vec{p}_{0}),\\\nonumber
[\hat{\textbf{b}}(\vec{p}),\hat{\textbf{b}}^{\dagger}(\vec{p}_{0})]=\frac{1}{(m_{1}^{2}-m_{2}^{2})}\delta(\vec{p}
-\vec{p}_{0}).
\end{eqnarray}
The above considerations enable us to determine commutators that
include the reality conditions and to discard the Hermitian
condition for the higher order derivative field. This doesn't
generate ghosts or negative norm states (\ref{calbd}) and these new
operators behave with positive norm states. They will enable to
include interaction potentials in an adequate way. However, the
Hermitian condition is lost for the complex higher order derivative
field, but it is recovered for the components of the field.

With these  tools, it will be possible to determine the Hamiltonian density (\ref{hamiltonparcial})
in terms of these annihilation an creation operators that include the reality conditions.

\subsection{Hamiltonian in Terms of Creation and Annihilation Ope\-rators}\label{denhamop}
In the preceding section we calculated commutator for  annihilation
and creation operators and established the basis of our method. Here
we applied these tools in order to calculate the Hamiltonian density
for the complex Bernard-Duncan using annihilation and creation
operators showing that the energy is bounded from below. The Hamiltonian
density (\ref{cons}) can be written in terms of
annihilation and creation operators. Term by term  the complex
Bernard-Duncan Hamiltonian density (\ref{cons}) can be pieced
together in order to obtain the Hamiltonian
\begin{eqnarray}\label{hamcrean}
H_{BDCA}=\int\!d^{3}p (m_{1}^{2}-m_{2}^{2})
( \frac{E_{1}}{2}[\hat{\textbf{a}}^{\dagger}(\vec{p})\hat{\textbf{a}}(\vec{p})+
\hat{\textbf{a}}(\vec{p})\hat{\textbf{a}}^{\dagger}(\vec{p})]
\\\nonumber
+\frac{E_{2}}{2} [\hat{\textbf{b}}^{\dagger}(\vec{p})\hat{\textbf{b}}(\vec{p})+
\hat{\textbf{b}}(\vec{p})\hat{\textbf{b}}^{\dagger}(\vec{p})]).
\end{eqnarray}
The density (\ref{hamcrean}) is real, Hermitian and positive
defined. It is a consequence to require that the solution
(\ref{phi}) for the equation (\ref{ecuacion}) satisfy the reality
conditions (\ref{realidaduno}) with the result (\ref{phidos}) if we
want Hermitian fields.

Thus, applying commutators (\ref{conmutacion}) on the Hamiltonian density we get
\begin{eqnarray}\label{hamiltonpua}
 H_{BDCA}=\int\!d^{3}p \left[ (m_{1}^{2}-m_{2}^{2}) \left\lbrace E_{1}
 \hat{\textbf{a}}^{\dagger}(\vec{p})\hat{\textbf{a}}(\vec{p})
+E_{2}\hat{\textbf{b}}^{\dagger}(\vec{p})\hat{\textbf{b}}(\vec{p})\right\rbrace\right.\\\nonumber
\left.+\frac{(E_{1}+E_{2})}{2}\delta(0)\right].
\end{eqnarray}
The above expression is bounded from below, a Hermitian quantity and it was
gotten with  annihilation and creation operators using reality conditions.
The expression (\ref{hamcrean})  will help us to
find a relationship between these annihilation and creation operators
and the operators for a real Klein-Gordon theory.

\subsection{Relationship between the Complex Bernard-Duncan Model and two
Real Klein-Gordon Fields} In preceding sections we introduced a
complex canonical transformation (\ref{realidad}) and concluded that
in order to quantize the theory in a right way is important to
introduce in the higher order derivative fields the reality
conditions instead of the Hermitian condition. In this form from the
complex Bernard-Duncan theory by means of reality conditions, we
obtain real fields and a real Hamiltonian density.

Using the complex canonical transformation (\ref{realidad}) we
obtain a very clear mapping where reality conditions are implicit,
but it doesn't define a way to introduce the interaction potentials.
However an alternative form which will allow us to find those is to
use reality conditions, although both of these theories differ by a
contact transformation (\ref{contact}).

The above statement will be demonstrated using the similarity
between (\ref{hamcrean}) and the real Klein-Gordon Hamiltonian
density (\ref{hamiltonkg}).

This suggest that the relationship between the annihilation and creation operators is
given by
\begin{eqnarray}\label{trans}
 \hat{\textbf{A}}(\vec{p})=(m_{1}^{2}-m_{2}^{2})^{\frac{1}{2}}\hat{\textbf{a}}(\vec{p}),\qquad
 \hat{\textbf{A}}^{\dagger}(\vec{p})=(m_{1}^{2}-m_{2}^{2})^{\frac{1}{2}}\hat{\textbf{a}}^{\dagger}(\vec{p}), \nonumber\\
 \hat{\textbf{B}}(\vec{p})=(m_{1}^{2}-m_{2}^{2})^{\frac{1}{2}}\hat{\textbf{b}}(\vec{p}),\qquad
 \hat{\textbf{B}}^{\dagger}(\vec{p})=(m_{1}^{2}-m_{2}^{2})^{\frac{1}{2}}\hat{\textbf{b}}^{\dagger}(\vec{p}).
\end{eqnarray}
By other hand this can be obtained using from (\ref{phidos}) to
(\ref{momentopiuno}) and analyzing the commutators for the
annihilation and creation operators (\ref{conmutacion}).

Using the transformation (\ref{trans}) we can map from the Hamiltonian density (\ref{hamcrean}) to
the Hamiltonian density of two Klein-Gordon fields (\ref{hamiltonkg}) in terms of
annihilation and creation operators $\hat{\textbf{A}},\hat{\textbf{B}}$.

Invoking the commutators (\ref{conmutacion}) and the contact
transformation (\ref{trans}), we obtain the desired commutators
\begin{eqnarray}
 [\hat{\textbf{A}}(\vec{p}),\hat{\textbf{A}}^{\dagger}(\vec{p}_{0})]=\delta(\vec{p}
-\vec{p}_{0}),\qquad
[\hat{\textbf{B}}(\vec{p}),\hat{\textbf{B}}^{\dagger}(\vec{p}_{0})]=\delta(\vec{p}
-\vec{p}_{0}),
\end{eqnarray}
that are the usual ones.

The annihilation and creation operators for the Klein-Gordon Hamiltonian density (\ref{hamiltonkg})
are linked to the fields in a very usual way
\begin{eqnarray}\label{amayuscula}
 \psi_{1}=\int\!\frac{d^{3}p}{(2\pi)^{\frac{3}{2}}(2E_{1})^{\frac{1}{2}}}(\hat{\textbf{A}} e^{-iE_{1}t+i\vec{p}\cdotp\vec{x}}
+\hat{\textbf{A}}^{\dagger} e^{iE_{1}t-i\vec{p}\cdotp\vec{x}}),\nonumber\\
\psi_{2}=\int\!\frac{d^{3}p}{(2\pi)^{\frac{3}{2}}(2E_{2})^{\frac{1}{2}}}(\hat{\textbf{B}} e^{-iE_{2}t+i\vec{p}\cdotp\vec{x}}
+\hat{\textbf{B}}^{\dagger} e^{iE_{2}t-i\vec{p}\cdotp\vec{x}}).
\end{eqnarray}
It has a kinship given by (\ref{contact}) where we saw that these
fields are the components from a complex field except by a contact
transformation. So, if we consider as starting point that the higher
order derivative theory is complex and using the projection of the
reality conditions, we finish with a real theory. Must be emphasized
that if we restrict our original theory to be an Hermitian theory
this mapping can't be done. Now, in view of this we don't have a
clear way in order to aggregate interaction potentials. The complex
canonical transformation fixes a mapping and establish the reality
conditions, but it can't provide these potentials. However, this
transformation gives the reality conditions and we will show that by
means of these conditions, we can establish the interaction
potentials.

It is instructive to summarize the above method. Using the complex
Bernard-Duncan theory described by its Hamiltonian density, its
complex phase space $(\phi,\dot{\phi},\pi_{\phi},\pi_{\dot{\phi}})$
and its annihilation and creation operators is possible to apply the
reality conditions, which reduce the grades of freedom from 8 to 4,
instead of the Hermitian conditions. However, these non-Hermitian
conditions generate a phase space where their fields
$(\psi_{1},\psi_{2},\pi_{\psi_{1}},\pi_{\psi_{2}})$ are Hermitian.

\section{An Interaction Potential in The Complex Ber\-nard Duncan Model}\label{interaction}
All the systems considered in the previous sections were composed of
non-interacting entities. For a real contact between the theory and
experiment, one must take into account the interparticle
interactions operating in the system. As stated early, Pais and
Uhlenbeck established a method in order to face the higher order
derivative theories which Hawking \cite{Cer} describe briefly.

In this usual formulation is common to suppose an Hermitian higher
order derivative field then to apply in the higher order derivative
Lagrangian formulation a non-local transformation, in order to
obtain a  Lagrangian density of two real Klein-Gordon fields which
differ by a sign \cite{Pai}. A possible interparticle potential
would be obtained by means of supposing the higher order derivative
field into Lagrangian density  behave similar to a real Klein-Gordon
field so, the interaction potential given by $\varphi^{4}$ would
generate an energy interchange.

Using the non-local transformation \cite{Cer} for $\varphi^{4}$, we obtain the effective
potential
\begin{equation}\label{pnolocal}
 V=m_{1}^{2}\chi_{1}^{2}-m_{2}^{2}\chi^{2}_{2}+\frac{4\lambda}
{(m_{2}^{2}-m_{1}^{2})^{2}}
(\chi_{1}-\chi_{2})^{4}.
\end{equation}
This potential  has conflicts linked to the stability. From
a perturbative deve\-lopment, since it isn't bounded below and  doesn't have a lower energy state.

\begin{figure}[ht]
\centering
  \includegraphics[scale=0.20]{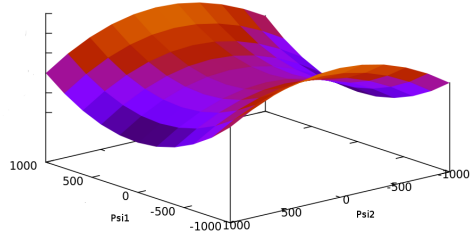}
\caption{Anomalous interaction potential derived from the non-local
transformation in \cite{Cer}.}\label{hawking}
\end{figure}

The graphic for the potential (\ref{pnolocal}) is given by the figure \ref{hawking} and it shows an inflection point
which isn't stable resulting impractical to do a perturbative method around this point.
Now, we can return to the formulation used in this work.
Here we develop a mechanism in order to attach interaction potentials into the
complex Bernard-Duncan model using the reality conditions.

\begin{figure}[ht]
\centering
  \includegraphics[scale=0.50]{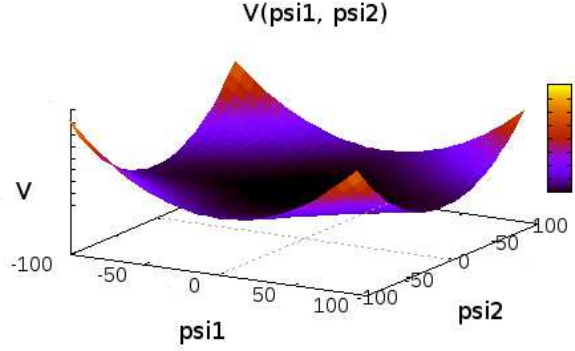}
\caption{Potential $V\mid_{RC}$ that has a perturbative
expansion.}\label{kgco}
\end{figure}

The procedure is based specifically, i.e. on the real reduction, we
consider all the possible interactions in the complex theory that
are real quantities applying the reality conditions (\ref{reali}).
The simplest examples are
\begin{eqnarray}\label{potuno}
  V^{1}=\int\!d^{3}x\frac{g_{1}}{4!(m_{1}^{2}-m_{2}^{2})^{2}}[m_{1}^{2}\phi-\pi_{\dot{\phi}}]^{4},\\
V^{2}=\int\!d^{3}x\frac{g_{2}}{4!(m_{1}^{2}-m_{2}^{2})^{2}}[m_{2}^{2}\phi-\pi_{\dot{\phi}}]^{4}
 \end{eqnarray}
and
\begin{equation}\label{potdos}
 V^{3}=\int\!d^{3}x\frac{g_{3}}{4!(m_{1}^{2}-m_{2}^{2})^{2}}[m_{1}^{2}\phi-\pi_{\dot{\phi}}]^{2}
[m_{2}^{2}\phi-\pi_{\dot{\phi}}]^{2}.
\end{equation}
These expressions in terms of the fields $\psi_{1}$, $\psi_{2}$
including the reality conditions (\ref{reali})  are
\begin{eqnarray}\label{unoatres}
 V^{1}\mid_{RC}=\int\!d^{3}x\frac{g_{1}}{4!}\psi_{2}^{4},
\qquad V^{2}\mid_{RC}=\int\!d^{3}x
\frac{g_{2}}{4!}\psi_{1}^{4},\\
V^{3}\mid_{RC}=\int\!d^{3}x\frac{g_{3}}{4!}\psi_{2}^{2}\psi_{1}^{2}\nonumber
\end{eqnarray}
and we can join these potentials to obtain the effective potential
\begin{equation}
 V\mid_{RC}=m_{1}^{2}\psi_{1}^{2}+m_{2}^{2}\psi_{2}^{2}+V^{1}\mid_{RC}+V^{2}\mid_{RC}+V^{3}
\mid_{RC},
\end{equation}
with its graphic given by the figure \ref{kgco}.
This shows that we have achieved to obtain a potential with a minimum which we can
apply a perturbative method.

In (\ref{hamiltonkg}), the fields $\psi_{1}$ and $\psi_{2}$ obey  free
equations of motion. Then by means of this free description we can write the S-matrix.

 The S-matrix elements are
\begin{equation}
S_{fi}= _{out}<k'_{1},k'_{2},...|T\exp[-i\int_{-\infty}^{\infty}dt\ V^{k}\mid_{RC}(\psi_{1},\psi_{2})\mid_{cre}]|k_{1},k_{2},...>_{in}
\end{equation}
with $k=1,2,3$. It is worth to mention that the reality conditions
produce Hermitian fields $\psi$ then we wrote it in an usual way. So
we can use (\ref{amayuscula}) in order to do a perturbative
expansion. From this description the Wick's theorem can be shown and
the time-order product is reduced into the normal ordered product as
we usually do. However, we pay attention to irreducible diagrams
$\frac{1}{i}\Sigma_{j}(p)$ with $j=1,2$ in order to describe self
energy process. The propagator is
\begin{equation}\label{gr}
G^{(2)}_{cj}(p)=\frac{i}{p^{2}-m_{Bj}^{2}-\Sigma_{j}(p)},
\end{equation}
or
\begin{equation}
[G^{(2)}_{cj}(p)]^{-1}=G_{0j}(p)^{-1}-\frac{1}{i}\Sigma_{j}(p).
\end{equation}
It is important to consider the relationship between the physical mass and the complete propagator
\begin{equation}
G^{(2)}_{cj}(p)=\frac{i}{p^{2}-m_{physj}^{2}}
\end{equation}
and to consider
\begin{equation}
m_{physj}^{2}=m_{Bj}^{2}+\Sigma_{j}(p).
\end{equation}
Now, it is necessary to include a new definition in order to obtain the two points vertex function defined by
\begin{equation}
G_{cj}^{(2)}(p)\Gamma_{j}(p)=i,
\end{equation}
which is finally
\begin{equation}
\Gamma_{j}(p)=p^{2}-m_{Bj}^{2}-\Sigma_{j}(p).
\end{equation}
Using the above expression for each mass
\begin{eqnarray}
\Gamma_{2}(p)=p^{2}-m_{B2}^{2}-\Sigma_{2}(p),\qquad \Gamma_{1}(p)=p^{2}-m_{B1}^{2}-\Sigma_{1}(p),
\end{eqnarray}
we have obtained the two points vertex functions.

\begin{figure}[ht]
\centering
  \includegraphics[scale=0.10]{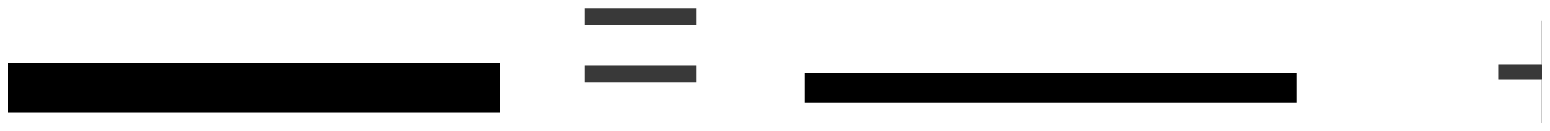}
  \includegraphics[scale=0.10]{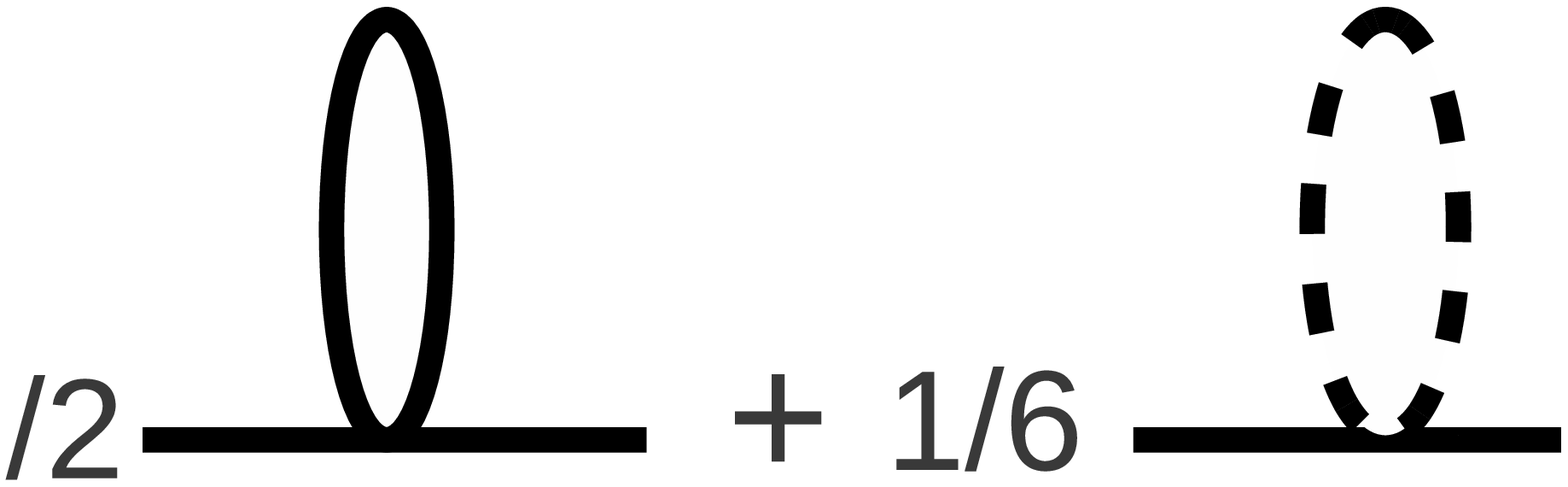}\\
  \includegraphics[scale=0.12]{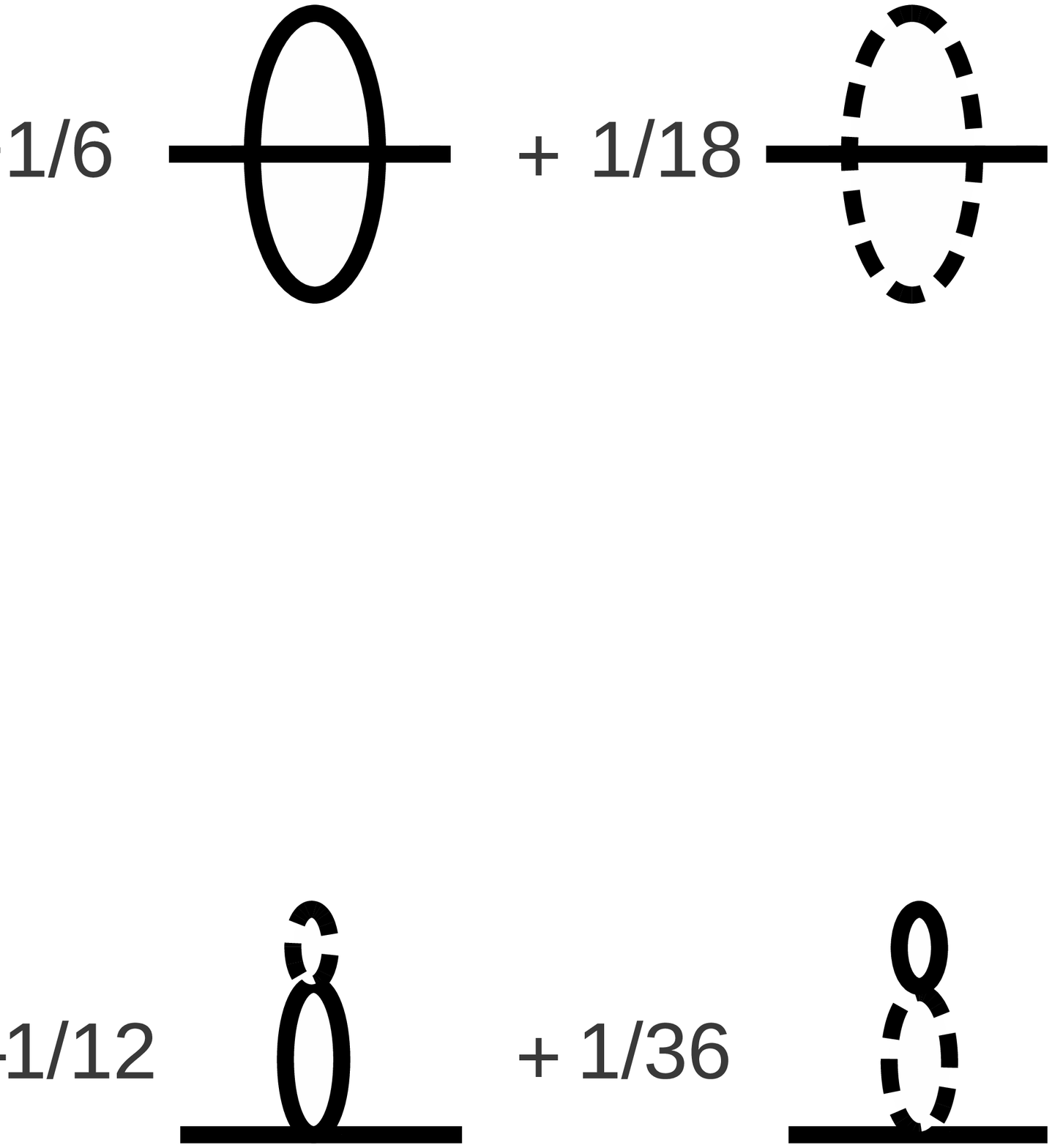}
  \includegraphics[scale=0.12]{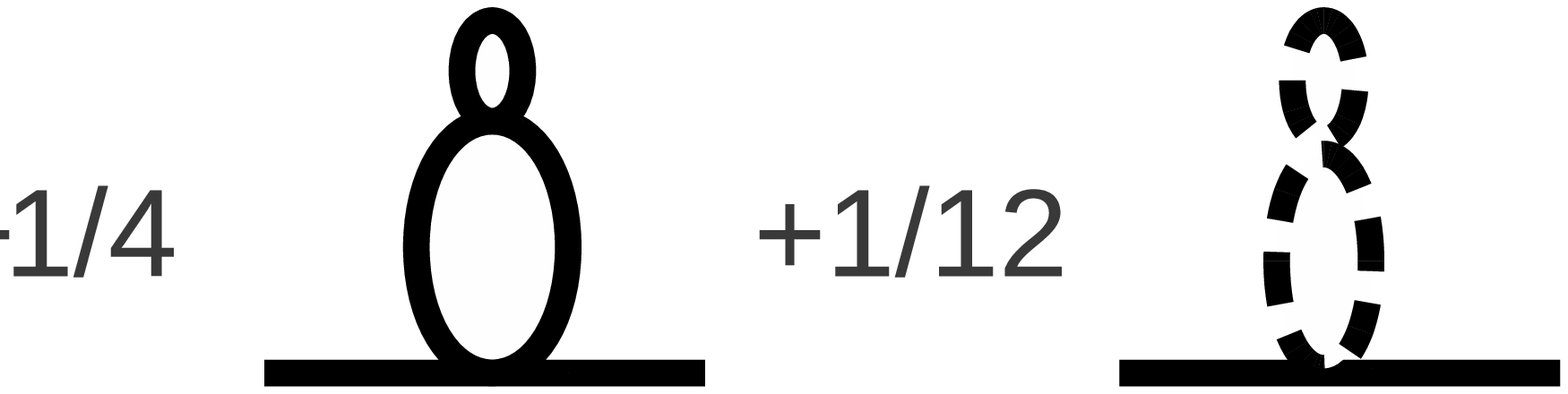}
\caption{Two legs diagrams for the compleat propagator
$G^{(2)}_{c2}(p)$  until order
$\mathcal{O}(g^{3}_{1,3})$.}\label{regularizacionuno}
\end{figure}

According to the Feynman rules, we fix the propagators by means of
including self-interactions. For the propagator associated with
$m_{B2}$, we obtain the figure \ref{regularizacionuno}.
The analytic expression for $\Sigma_{2}(p)$ is
\begin{eqnarray}\label{patasuno}
\Sigma_{2}(p)=\frac{g_{1}}{2(2\pi)^{4}}
\int\frac{d^{4}k_{E}}{k_{E}^{2}+m^{2}_{B2}}+\frac{g_{3}}{6(2\pi)^{4}}\int\frac{d^{4}k_{E}}{k_{E}^{2}+m^{2}_{B1}}\\
-\frac{g_{1}^{2}}{6(2\pi)^{8}}
\int \frac{ d^{4}p_{E1}
d^{4}p_{E2}}{[(p_{E1}+p_{E2}+q_{E})^{2}+m_{B2}^{2}][p_{E1}^{2}+m_{B2}^{2}][p_{E2}^{2}
+m^{2}_{B2}]}\nonumber\\
-\frac{g_{3}^{2}}{18(2\pi)^{8}}
\int \frac{d^{4}p_{E1}
d^{4}p_{E2}}{[(p_{E1}+p_{E2}+q_{E})^{2}+m_{B1}^{2}][p_{E1}^{2}+m_{B1}^{2}][p_{E2}^{2}
+m^{2}_{B2}]}\nonumber\\
-\frac{g^{2}_{1}}{4(2\pi)^{8}}\int \frac{ d^{4}k_{E1}
d^{4}k_{E2}}{[k_{E1}^{2}+m_{B2}^{2}]^{2}[k_{E2}^{2}+m^{2}_{B2}]}-\frac{g_{2}g_{3}}{12(2\pi)^{8}}\nonumber\\
\int \frac{ d^{4}k_{E1}
d^{4}k_{E2}}{[k_{E1}^{2}+m_{B1}^{2}]^{2}[k_{E2}^{2}+m^{2}_{B1}]}
-\frac{g_{1}g_{3}}{12(2\pi)^{8}}\int \frac{d^{4}k_{E1}
d^{4}k_{E2}}{[k_{E1}^{2}+m_{B2}^{2}]^{2}[k_{E2}^{2}+m^{2}_{B1}]}\nonumber\\
-\frac{g_{3}^{2}}{36(2\pi)^{8}}\int \frac{d^{4}k_{E1}d^{4}k_{E2}}{[k_{E1}^{2}+m_{B1}^{2}]^{2}[k_{E2}^{2}+m_{B2}^{2}]}\nonumber
\end{eqnarray}
where $q_{E}=-p_{E}$. For the propagator associated with $m_{B1}$ we
obtain similar expressions and diagrams, but we have to interchange
dashed  lines to continuous lines and continuous lines to dashed
lines in the figure \ref{regularizacionuno}.

The analytic expression for $\Sigma_{1}(p)$ can be obtain changing
$g_{1}\rightarrow g_{2}$, $m_{B2}\rightarrow m_{B1}$ and
$m_{B1}\rightarrow m_{B2}$.

In the above expression have been developed a Wick rotation, because
it permit to separate the divergent part of integrals by means of
dimensional regularization.

The regularized expression for (\ref{patasuno}) is
\begin{eqnarray}
\Sigma_{2}(p)\approx
\frac{m^{2}_{2B}g_{1}}{2(4\pi)^{2}}[\frac{2}{\epsilon}+\Psi(2)+ln(\frac{4\pi\mu^{2}_{R}}{m^{2}_{2B}})]\\
+
\frac{m^{2}_{1B}g_{3}}{6(4\pi)^{2}}[\frac{2}{\epsilon}+\Psi(2)+ln(\frac{4\pi\mu^{2}_{IR}}{m^{2}_{1B}})]+\frac{g_{1}^{2}}{6(4\pi)^{4}}\times \nonumber\\
\left\lbrace 3m_{2B}^{2}[\frac{2}{\epsilon^{2}}+\frac{2}{\epsilon}(\frac{3}{2}+\Psi_{1}+log(\frac{4\pi\mu_{R}^{2}}{m_{2B}^{2}}))]+\frac{q^{2}}{2\epsilon}\right\rbrace \nonumber\\
+\frac{g_{3}^{2}m_{1B}^{2}}{18(4\pi)^{4}}\times     [ \frac{(4+\frac{2m_{2B}^{2}}{m_{1B}^{2}})}{\epsilon^{2}}+\frac{4}{\epsilon}(-\gamma+log(\frac{4\pi\mu_{RI}^{2}}{m_{1B}^{2}}))\nonumber\\
+ \frac{2m_{2B}^{2}}{m_{1B}^{2}\epsilon}[-\gamma+log(\frac{4\pi\mu_{RI}^{2}}{m_{2B}^{2}})]
+\frac{2}{\epsilon}+\frac{m_{2B}^{2}}{m_{1B}^{2}\epsilon}+\frac{q^{2}}
{2\epsilon m_{1B}^{2}} ]\nonumber\\
+\frac{g_{1}^{2}m_{2B}^{2}}{4(4\pi)^{4}}\times[\frac{4}{\epsilon^{2}}+\frac{2(\psi(1)+\psi(2))}{\epsilon}-\frac{4}{\epsilon}log(\frac{m_{2B}^{2}}{4\pi\mu_{R}^{2}})]
\nonumber\\
+\frac{g_{2}g_{3}m_{1B}^{2}}{12(4\pi)^{4}}[\frac{4}{\epsilon^{2}}+\frac{2(\psi(1)+\psi(2))}{\epsilon}-\frac{2}{\epsilon}log(\frac{m_{1B}^{2}}{4\pi\mu_{I}\mu_{RI}})]
\nonumber\\
+\frac{g_{3}g_{1}m_{1B}^{2}}{12(4\pi)^{4}}[\frac{4}{\epsilon^{2}}+\frac{2(\psi(1)+\psi(2))}{\epsilon}-\frac{2}{\epsilon}log(\frac{m_{2B}^{2}}{4\pi\mu_{R}\mu_{RI}})]
\nonumber\\
+\frac{g^{2}_{3}m_{2B}^{2}}{36(4\pi)^{4}}[\frac{4}{\epsilon^{2}}+\frac{2(\psi(1)+\psi(2))}{\epsilon}-\frac{2}{\epsilon}log(\frac{m_{1B}^{2}}{4\pi\mu_{RI}^{2}})]\nonumber
\end{eqnarray}
with $\Psi(2)=1-\gamma$, $\Psi(z)=\frac{\Gamma'(z)}{\Gamma(z)}$  and $\gamma$ being the Euler's constant.

\begin{figure}[ht]
\centering
  \includegraphics[scale=0.15]{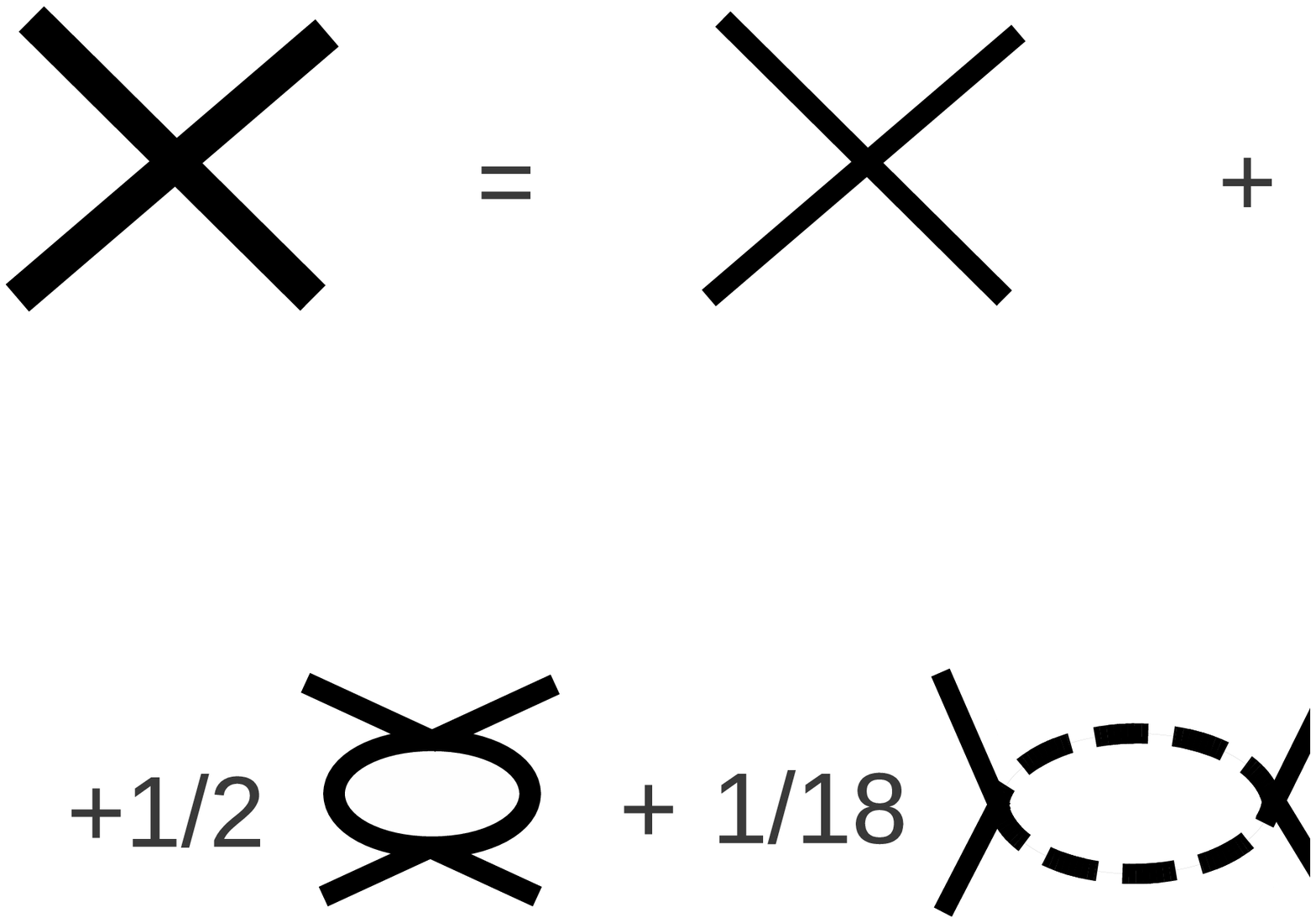}
  \includegraphics[scale=0.15]{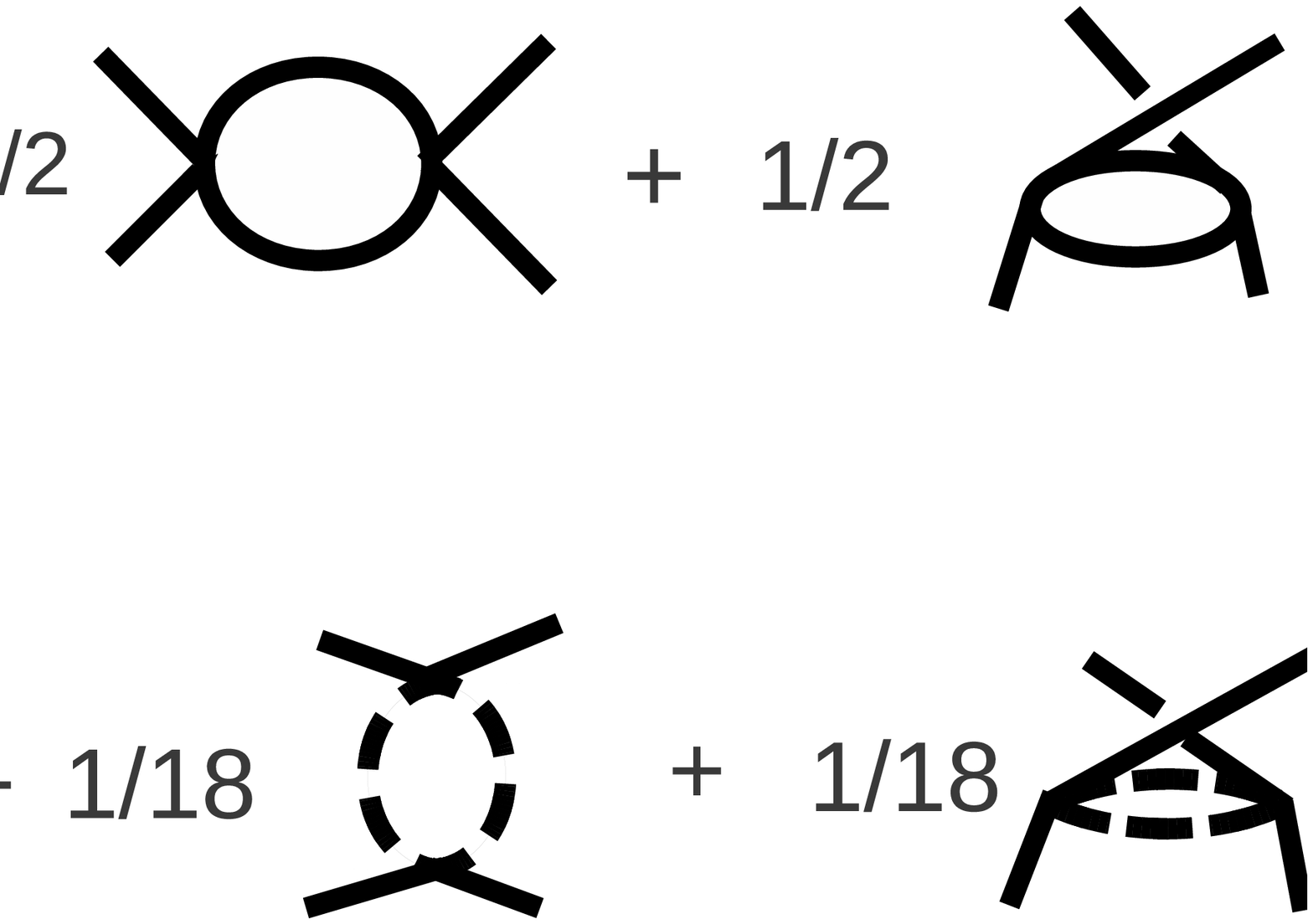}
\caption{One-loop four-points function for  $g_{1}$, $g_{3}$.}\label{dosamputados}
\end{figure}

The physical masses are
\begin{equation}
 m_{2phy}^{2}=-\Gamma_{2}^{(2)}(0),\qquad m_{1phy}^{2}=-\Gamma_{1}^{(2)}(0)
\end{equation}
and the relation between bare and renormalized masses is
\begin{eqnarray}
 m_{2B}^{2}=[1-\frac{1}{\epsilon}(\frac{g_{1}}{(4\pi)^{2}}+\frac{g_{1}^{2}}{2(4\pi)^{4}}
+\frac{g_{1}^{2}(\Psi(1)+\Psi(2))}{(4\pi)^{4}}\\
+\frac{g_{3}^{2}(\Psi(1)+\Psi(2))}{18(4\pi)^{4}})-(\frac{2g_{1}^{2}}{(4\pi)^{4}}-\frac{g_{3}^{2}}{9(4\pi)^{4}}+\frac{2g_{1}g_{2}}{(4\pi)^{4}})\frac{1}{\epsilon^{2}}]m_{2phy}^{2}
\nonumber\\
-[\frac{1}{\epsilon}(\frac{g_{3}}{(4\pi)^{2}}-\frac{8\gamma g_{3}^{2}}{9(4\pi)^{4}}
+\frac{g_{2}g_{3}(\Psi(1)+\Psi(2))}{6(4\pi)^{4}}\nonumber\\
+\frac{g_{1}g_{3}(\Psi(1)+\Psi(2))}{6(4\pi)^{4}})+(\frac{2g_{3}^{2}}{9(4\pi)^{4}}+\frac{4g_{1}g_{3}}{3(4\pi)^{4}}+\frac{4g_{2}g_{3}}{3(4\pi)^{4}})\frac{1}{\epsilon^{2}}]m^{2}_{1ph}
\nonumber
\end{eqnarray}
for $m^{2}_{1}$, we obtain
\begin{eqnarray}
 m_{1B}^{2}=[1-\frac{1}{\epsilon}(\frac{g_{2}}{(4\pi)^{2}}+\frac{3g_{2}^{2}}{2(4\pi)^{4}}
+\frac{2g_{2}^{2}(\Psi(1)+\Psi(2))}{(4\pi)^{4}}\\
+\frac{g_{3}^{2}(\Psi(1)+\Psi(2))}{18(4\pi)^{4}})-(\frac{g_{1}^{2}+2g_{2}^{2}}{(4\pi)^{4}}-\frac{2g_{3}^{2}}{9(4\pi)^{4}}+\frac{5g_{1}g_{2}}{4(4\pi)^{4}})\frac{1}{\epsilon^{2}}]m_{1phy}^{2}\nonumber\\
-[\frac{1}{\epsilon}(\frac{g_{3}}{3(4\pi)^{2}}-\frac{4\gamma g_{3}^{2}}{9(4\pi)^{4}}
+\frac{g_{1}g_{3}(\Psi(1)+\Psi(2))}{6(4\pi)^{4}}\nonumber\\
+\frac{g_{2}g_{3}(\Psi(1)+\Psi(2))}{6(4\pi)^{4}})+(\frac{2g_{3}^{2}}{9(4\pi)^{4}}+\frac{2g_{1}g_{3}}{3(4\pi)^{4}}+\frac{2g_{2}g_{3}}{3(4\pi)^{4}})\frac{1}{\epsilon^{2}}]m^{2}_{2ph}.
\nonumber
\end{eqnarray}
We now apply a similar treatment to $\Gamma^{(4)}$ in order to
renormalize the coupling
 constant $g_{j}$.
We can consider the new momenta $k_{E}=p_{4E}-p_{2E}=p_{1E}-p_{3E}$,
$k'_{E}=p_{3E}-p_{2E}$ and $k''_{E}=p_{1E}+p_{2E}=p_{3E}+p_{4E}$ in
order to calculate the four-points function to take into account the
amputated diagrams that are in the figure \ref{dosamputados}. The
analytic expression of the figure \ref{dosamputados} is
\begin{eqnarray}\label{expres}
 \Gamma_{1}^{(4)}(p_{1E}, p_{2E}, p_{3E}, p_{4E})=-ig_{1}\\
+\frac{ig_{1}^{2}}{2(2\pi)^{4}}\int\frac{d^{4}q_{E}}{[(q_{E}+k_{E})^{2}+m_{2B}^{2}][q_{E}^{2}+m_{2B}^{2}]}\nonumber\\
+\frac{ig_{1}^{2}}{2(2\pi)^{4}}\int\frac{d^{4}q_{E}}{[(q_{E}+k'_{E})^{2}+m_{2B}^{2}][q_{E}^{2}+m_{2B}^{2}]}\nonumber\\
+\frac{ig_{1}^{2}}{2(2\pi)^{4}}\int\frac{d^{4}q_{E}}{[(q_{E}+k''_{E})^{2}+m_{2B}^{2}][q_{E}^{2}+m_{2B}^{2}]}\nonumber\\
+\frac{ig_{3}^{2}}{18(2\pi)^{4}}\int\frac{d^{4}q_{E}}{[(q_{E}+k_{E})^{2}+m_{1B}^{2}][q_{E}^{2}+m_{1B}^{2}]}\nonumber\\
+\frac{ig_{3}^{2}}{18(2\pi)^{4}}\int\frac{d^{4}q_{E}}{[(q_{E}+k''_{E})^{2}+m_{1B}^{2}][q_{E}^{2}+m_{1B}^{2}]}\nonumber\\
+\frac{ig_{3}^{2}}{18(2\pi)^{4}}\int\frac{d^{4}q_{E}}{[(q_{E}+k'_{E})^{2}+m_{1B}^{2}][q_{E}^{2}+m_{1B}^{2}]}.\nonumber
\end{eqnarray}

Putting
\begin{eqnarray}
 F(s,m,\mu)\equiv\int_{0}^{1}\!dx \ \ln(\frac{sx(1-x)+m^{2}}{4\pi\mu^{2}})
\end{eqnarray}
the regularized expression for (\ref{expres}) is
\begin{eqnarray}
\Gamma_{1}^{(4)}(p_{1E}, p_{2E}, p_{3E}, p_{4E})\approx
-ig_{B1}\mu^{\epsilon}_{R}+\frac{3ig_{B1}^{2}\mu_{R}^{\epsilon}
}{16\pi^{2}\epsilon}\\
+\frac{ig_{B3}^{2}\mu_{RI}^{\epsilon}
}{48\pi^{2}\epsilon}
-\frac{ig_{B1}^{2}\mu_{R}^{\epsilon}}{32\pi^{2}}
[3\gamma+F(k_{E}^{2},m_{2},\mu_{R})
\nonumber\\
+F(k_{E}^{'2},m_{2},\mu_{R})
+F(k_{E}^{''2},m_{2},\mu_{R})]\nonumber\\
-\frac{ig_{B3}^{2}\mu_{RI}^{\epsilon}}{288\pi^{2}}
[3\gamma+F(k_{E}^{2},m_{1},\mu_{RI})\nonumber\\
+F(k_{E}^{'2},m_{1},\mu_{RI})
+F(k_{E}^{''2},m_{1},\mu_{RI})].\nonumber
\end{eqnarray}
As in $\Gamma^{(2)}_{1}$, we obtain similar expressions to
$\Gamma^{(4)}_{2}$ interchanging dashed  lines to conti\-nuous lines
and continuous lines to dashed lines. For the analytic expression it
is changed $g_{1}\rightarrow g_{2}$, $m_{1}\rightarrow m_{2}$ and
$m_{2}\rightarrow m_{1}$.
\begin{figure}[ht]
\centering
  \includegraphics[scale=0.15]{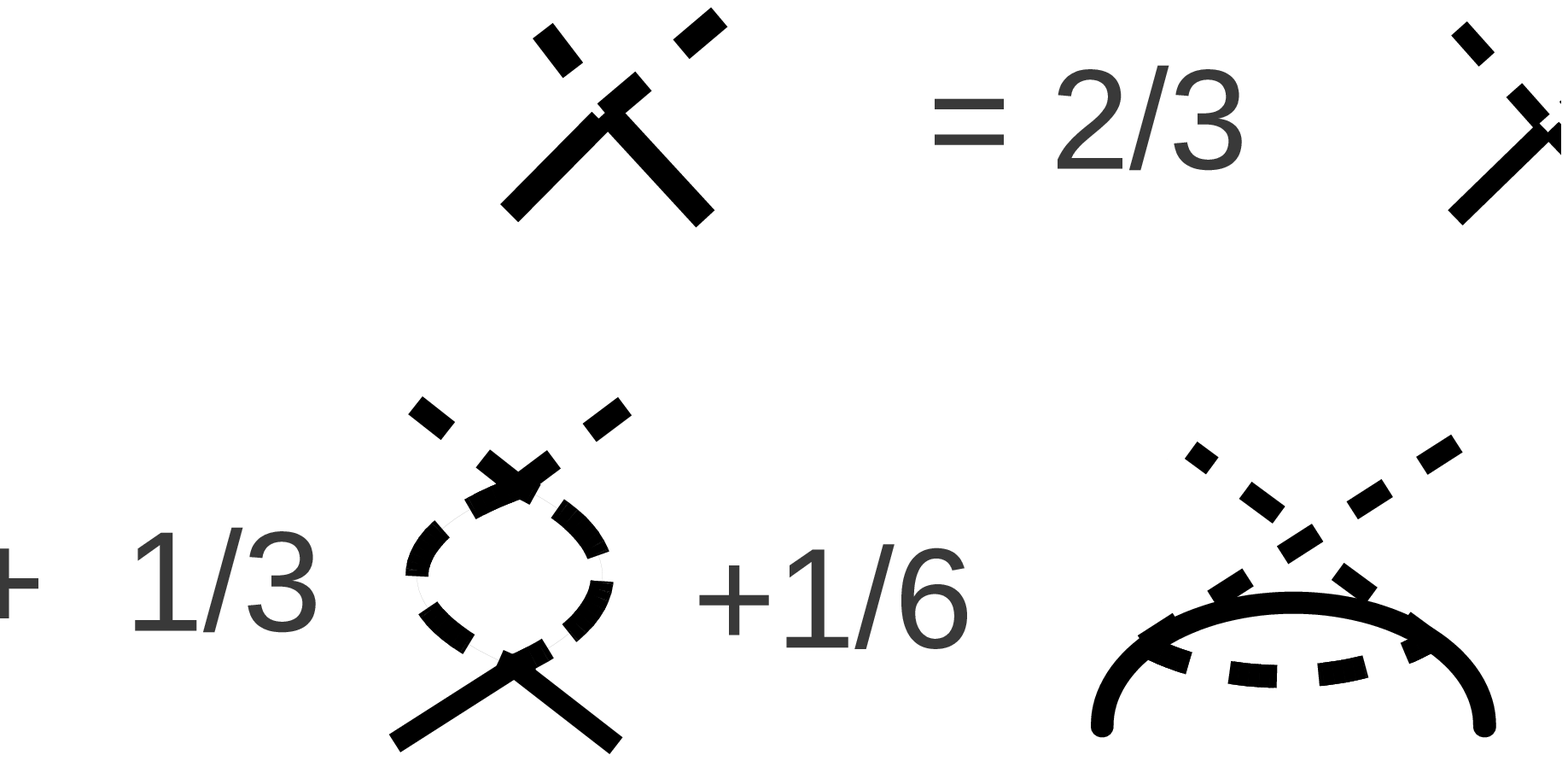}
  \includegraphics[scale=0.15]{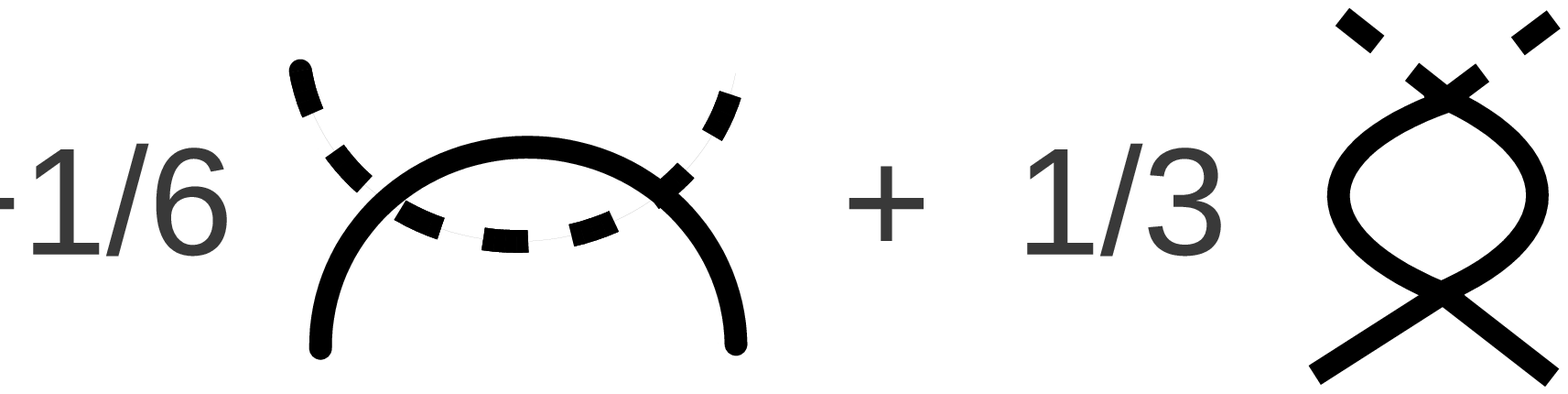}
\caption{One-loop four-points function with different external legs for
 $g_{1}$, $g_{2}$ and $g_{3}$.}\label{incampos}
\end{figure}
with $j=1,2,3$ and $k=1,2,3,4$. The corresponding renormalization is
\begin{eqnarray}
g_{B1}\approx
g_{1phy}\mu^{-\epsilon}_{R}+\frac{3g_{1phy}^{2}\mu_{R}^{-2\epsilon}
}{16\pi^{2}\epsilon}\\
+\frac{9g_{3phy}^{2}\mu_{RI}^{-\epsilon}\mu_{R}^{-\epsilon}
}{192\pi^{2}\epsilon}
-\frac{g_{1phy}^{2}\mu_{R}^{-\epsilon}}{32\pi^{2}}
[3\gamma+F(0,m_{2phy},\mu_{R})
\nonumber\\
+F(0,m_{2phy},\mu_{R})
+F(0,m_{2phy},\mu_{R})]\nonumber\\
-\frac{9g_{3phy}^{2}\mu_{RI}^{-\epsilon}}{1152\pi^{2}}
[3\gamma+F(0,m_{1phy},\mu_{RI})\nonumber\\
+F(0,m_{1phy},\mu_{RI})
+F(0,m_{1phy},\mu_{RI})]\nonumber
\end{eqnarray}
and for the other coupling constant, we have
\begin{eqnarray}
g_{B2}\approx
g_{2phy}\mu^{-\epsilon}_{I}+\frac{3g_{1phy}^{2}\mu_{I}^{-2\epsilon}
}{16\pi^{2}\epsilon}\\
+\frac{9g_{3phy}^{2}\mu_{RI}^{-\epsilon}\mu_{I}^{-\epsilon}
}{192\pi^{2}\epsilon}
-\frac{g_{1phy}^{2}\mu_{I}^{-\epsilon}}{32\pi^{2}}
[3\gamma+F(0,m_{1phy},\mu_{I})
\nonumber\\
+F(0,m_{1phy},\mu_{I})
+F(0,m_{1phy},\mu_{I})]\nonumber\\
-\frac{9g_{3phy}^{2}\mu_{RI}^{-\epsilon}}{1152\pi^{2}}
[3\gamma+F(0,m_{2phy},\mu_{RI})\nonumber\\
+F(0,m_{2phy},\mu_{RI})
+F(0,m_{2phy},\mu_{RI})].\nonumber
\end{eqnarray}
Because we have three vertexes the regularization for the coupling
constant $g_{3}$,  given by diagrams in figure \ref{incampos}, is

\begin{eqnarray}
g_{3}\approx\frac{3g_{3phy}\mu^{-\epsilon}_{RI}}{2}\\
+\frac{9g^{2}_{3phy}\mu_{RI}^{-2\epsilon}}{64\pi^{2}\epsilon}
+\frac{3g_{3phy}\mu_{RI}^{-\epsilon}(\mu_{R}^{-\epsilon}g_{1phy}+\mu_{I}^{-\epsilon}g_{2phy})}{2\pi^{2}\epsilon}
-\frac{3g^{2}_{3phy}\mu_{RI}^{-\epsilon}}{192\pi^{2}} [2\gamma\nonumber\\
+\int_{0}^{1}\! \ln(\frac{m_{2phy}^{2}+
(m_{1phy}^{2}-m_{2phy}^{2})x}{4\pi\mu_{RI}^{2}})dx\nonumber\\
+\int_{0}^{1}\! \ln(\frac{m_{2phy}^{2}+
(m_{1phy}^{2}-m_{2phy}^{2})x}{4\pi\mu_{RI}^{2}})dx]\nonumber\\
-\frac{3g_{3phy}}{96\pi^{2}} [(\mu_{R}^{-\epsilon}g_{1phy}+\mu_{I}^{-\epsilon}g_{2phy})\gamma
+\mu_{R}^{-\epsilon}g_{1phy}F(0,m_{2},\mu_{R})\nonumber\\
+\mu_{I}^{-\epsilon}g_{2phy}F(0,m_{1},\mu_{I})].\nonumber
\end{eqnarray}
The above method works without problems, but we have eliminated the
Hermiticity condition for the higher order derivative field. It
allowed us to include  interaction potentials in the higher order
derivative theory that were mapped to real interaction potentials by
means of the reality conditions, see figure \ref{kgco}. The mapping
here described results in a renormalizable theory without inherent
pathologies from the higher order derivative theories \cite{Cer} as
it was shown in the figure \ref{hawking}.

\section{Conclusions}\label{conclu}
The higher order derivative theories are an alternative description
of the nature that can be encoded to the usual first order
mechanics. The above is seen by means of the existence of a
canonical transformation between the real Bernard-Duncan Hamiltonian
density and the Hamiltonian density of two real Klein-Gordon fields
with opposite sign \cite{Pai}. In the classical context the
equations of motion are equivalent and also to the quantum level but
in this case the system suffers irremediable inconsistencies once we
include interactions in the system. In this case the effective
potential is instable, see figure \ref{hawking}.

If we look at a canonical transformation from the Bernard-Duncan
Hamiltonian Density to Hamiltonian density of two Klein-Gordon
fields with positive sign, it is necessary to introduce a complex
canonical transformation \cite{dec} resulting in a complex extension
of the Bernard-Duncan Hamiltonian. Fundamental to this formulation
is to show that a restricted complex description is equivalent to
the usual classic description. This is achieved by showing that the
equations of motion resulting from both formulations are equivalent.

Since, our starting point is now a complex higher order derivative
theory to quantize the system we disregard the Hermiticity axiom.
The alternative to this condition is to use the so called reality
conditions \cite{ashtekar}. Using these conditions we map the
complex higher order derivative theory to a real first order theory
in Section 4. These conditions are really constraints to the complex
theory and reduce the degrees of freedom and the most important are
consistent with a class of interactions. In spite of the higher
order derivative field isn't Hermitian, if the reality conditions
are considered, the resulting fields and the Hamiltonian density
will be Hermitian quantities. The initial description isn't
Hermitian, but the reduction imposed by the reality condition is
Hermitian.

The way to attach the reality conditions in this work was through
annihilation and creation operators instead of considering the
Hermiticity conditions. The complex extension give us a greater
flexibility and in this way it is still possible to obtain an
Hermitian theory (\ref{hamiltonpua}).

The reality conditions also allow us to establish the interaction
potentials  (\ref{potuno})-(\ref{potdos}) that generate real
interaction  potentials (\ref{unoatres}) imposing the conditions
(\ref{corean}) and resulting a total potential that has a minimum
critical point. In consequence, it makes possible a perturbative
expansion as in the figure \ref{kgco} and  a regularizable and
renormalizable theory.

It is instructive here to summarize this work. According to section
\ref{combedu}, we introduced the complex Bernard-Duncan model using
the action and it was established the complex Bernard-Duncan
Hamiltonian density. Using this complex density, we set a mapping
from this one to the real Hamiltonian density of two Klein-Gordon
fields \cite{dec}. By means of this mapping, we obtained the reality
conditions that, according to section \ref{crea}, was applied by
means of annihilation and creation operators and they replaced the
Hermiticity conditions resulting the Hamiltonian density of two real
Klein-Gordon fields. This Hamiltonian density, according to section
\ref{interaction}, allow us to include interaction potentials in a
regularizable and renormalizable way.

\end{document}